\documentclass[aps,prb,twocolumn,floatfix,showpacs]{revtex4}
\usepackage{graphicx}
\usepackage{amsmath}
\usepackage{verbatim}

\newcommand{\bra}[1]{\left\langle {#1} \right|}
\newcommand{\ket}[1]{\left| {#1} \right\rangle}
\newcommand{\rbra}[1]{\left( {#1} \right|}
\newcommand{\rket}[1]{\left| {#1} \right)}
\newcommand{\vect}[1]{\boldsymbol{{#1}}}

\newlength{\fig}
\newlength{\smallfig}
\begin{document}

\title{Chiral Spin Textures of Strongly Interacting Particles in Quantum Dots}
\author{Catherine J. Stevenson}
\author{Jordan Kyriakidis}
\homepage{http://quantum.phys.dal.ca}
\affiliation{Department of Physics and Atmospheric Science, Dalhousie
  University, Halifax, Nova Scotia, Canada, B3H 3J5}

\begin{abstract}
  We probe for statistical and Coulomb induced spin textures among the
  low-lying states of repulsively-interacting particles confined to
  potentials that are both rotationally and time-reversal invariant.
  In particular, we focus on two-dimensional quantum dots and employ
  configuration-interaction techniques to directly compute the
  correlated many-body eigenstates of the system.  We produce spatial
  maps of the single-particle charge and spin density and verify the
  annular structure of the charge density and the rotational
  invariance of the spin field.  We further compute two-point spin
  correlations to determine the correlated structure of a single
  component of the spin vector field.  In addition, we compute
  three-point spin correlation functions to uncover chiral structures.
  We present evidence for both chiral and quasi-topological spin
  textures within energetically degenerate subspaces in the three- and
  four-particle system.  \\

\end{abstract}

\date{\today}


\maketitle

\section{Introduction}
\label{sec:intro}

The investigation of correlations among electrons confined to quantum
dots (QDs) is an active area of research in condensed matter physics
due to their experimental
tunability,\cite{kouwen01:few.elect.quant.dots,
  bukow02:quant.dot.resear} their theoretical
efficacy,\cite{reiman02:elect.struc.quant.dots} and their application
in, for example, quantum information
science.\cite{burkar06:theor.solid.state, engel04:contr.spin.qubit,
  Krenner05:Recent-advances, Taylor2005, Foletti2009, Laird2010} For
two-dimensional QDs with circular confinement, the combination of
confinement and long-range Coulomb repulsion results in charge
densities peaked in annular regions about the dot center.
Calculations of two-point correlation functions\cite{Ghosal2007} have
further revealed that these confined electrons exhibit textures akin
to Wigner molecules.  Numerical work has shown these behaviors to
depend on the size and shape of the quantum dot, as well as the
strength of the applied magnetic field.\cite{Ghosal2006}

Since the spin of confined electrons provides a viable implementation
of qubits,\cite{Loss98:Quantum-computation,Burkard99:Coupled-quantum}
an understanding of the configurations and correlations formed among
confined spins is crucial for the implementation of spin-qubits.
Furthermore, the degree of control in fabrication and manipulation of
QDs makes them ideal environments for the study of fundamental
behaviors of both spin and charge.  In this paper, we investigate the
correlations that exist between the spins of electrons trapped in
circular QDs.  We are specifically interested in the formation of
topological spin textures that may arise due to interaction or
statistical effects among the confined charges.

A significant challenge in the development of spin-based quantum dot
quantum computing is the suppression of decoherence of the spin-states
for time-scales much longer than the time required to controllably
flip a spin.\cite{Hanson2007, Ramsay2010} The potential to encode
information using topological degrees of freedom is appealing since
the enhanced stability can mitigate the burden of error-correction.
Various schemes which exploit a system's topological structure have
been proposed.\cite{Averin2001, Kitaev2003, Das-Sarma2005, Bombin2007}
The advantage of these is that they are \emph{physically}
fault-tolerant; they are immune to local perturbations that degrade
the coherent evolution of the state, a necessary ingredient in quantum
computation.\cite{DiVincenzo2000} Possible systems include
two-dimensional spin models\cite{Kitaev2003} (for example, atoms in an
optical lattice\cite{Duan2003}) and fractional quantum Hall
systems.\cite{Das-Sarma2005} These proposals rely on the existence of
non-Abelian anyons in the excitation spectrum of the models for the
information processing.

However, even in the absence of anyonic excitations, textures with
topological structure are expected to be long-lived
in-and-of-themselves due to their global correlations.  Since
correlations decrease decoherence,\cite{Bertoni2005, Climente2006,
  Climente2007} these topological structures could form an important
processing element in more conventional quantum computing schemes.
Even in finite-sized systems, where true topological stability likely
does not occur, the relevant relaxation and decoherence times can be
significantly enhanced.  In Ref.~\onlinecite{Bertoni2005}, for
example, even moderate charge correlations were sufficient to more
than double the decoherence time.

Numerical work has predicted the formation of spin textures in QDs
immersed in a magnetic field.\cite{Governale2002} Experimental
evidence suggests the existence of fermionic spin textures in a
two-dimensional electron gas (2DEG) confined in semiconductor
heterostructures,\cite{Kumada2006} in vertical QDs,\cite{Nishi2006}
and in few-electron lateral QDs.\cite{Ciorga2003, Sachrajda2004}

A topological spin qubit would be advantageous as it could be more
robust against local environmental perturbations.  Nuclear magnetic
resonance measurements of GaAs/AlGaAs quantum wells have shown
evidence for the localization of topological skyrmion spin textures as
the temperature approaches 0 K.\cite{Khandelwal2001, Gervais2005}
Recently, topological spin textures have been experimentally observed
in topological insulators by means of spin-resolved angle-resolved
photoemission spectroscopy.\cite{Hsieh2009} The emergence of a
skyrmion lattice has also been detected in the chiral magnet MnSi
using neutron scattering.\cite{Muehlbauer2009} Topological textures
that are predicted to appear in QDs include
vortices\cite{Saarikoski2004, Tavernier2004, Saarikoski2005, Yang2007,
  Anisimovas2008} and merons.\cite{Yang2005, Petkovic2007,
  Milovanovic2009} Vortices occur in the presence of a strong external
magnetic field, when electron current circulates in a plane around
localized regions of low electron density.  Merons are topological
spin textures characterized by a central ``up'' or ``down'' spin which
smoothly transitions into an in-plane 2$\pi$-winding along its
boundary.\cite{Moon1995} As developed in the theory, the realization
of both types of quasiparticles requires the presence of an external
magnetic field.

In this work, we present evidence for the existence of spin textures
in circular QDs for both three-electron and four-electron systems in
the absence of an external magnetic field.  The electronic wave
functions are calculated by configuration interaction techniques.
Two-point and three-point spin correlations are calculated in order to
uncover both correlation and chirality in the spin textures which are
concealed in superpositions of different configurations.

In Sec.~\ref{sec:system}, we introduce our model and the foundation
upon which our calculations are based.  Section~\ref{sec:spinops}
describes specifically the spin correlation calculations used to
examine the spin textures in the QD.  We then go on to describe our
results for systems of three (Sec.~\ref{N3}) and four (Sec.~\ref{N4})
interacting particles.  We conclude with a summary of our findings,
their implications, and suggestions for further investigations in
Sec.~\ref{sec:summary}.

\section{Quantum Dot System}
\label{sec:system}
Our system consists of $N$ interacting quasiparticles of charge $e$,
bound to a two-dimensional (2D) plane and laterally confined by a
parabolic potential.  The 2D Hamiltonian used to describe this
``standard model'' is
\begin{subequations}
\label{eq:hamil}
\begin{equation}
  \label{eq:intHam}
  \hat{\mathcal{H}}=\sum_i^N \hat{h}_i + \frac{1}{2}\sum_{i \neq j}^N
  \frac{e^2}{\epsilon | \hat{\vect{r}}_i-\hat{\vect{r}}_j |},
\end{equation}
where $\epsilon$ is the dielectric constant of the medium and
$\hat{h}$ is the single-particle Hamiltonian describing harmonic
confinement;
\begin{equation}
  \label{eq:nonIntHam}
  \hat{h} = \frac{1}{2m^*}
  \left(
    \hat{\vect{p}} + \frac{e}{c}\widehat{\vect{A}}
    \left(\hat{\vect{r}} \right)
  \right)^2 
  + \frac{1}{2} m^* \omega_0^2 \hat{r}^2,
\end{equation}
\end{subequations}
where $m^*$ is the effective mass, $\hat{\vect{r}} =
(\hat{x},\hat{y})$ the position operator, and $\omega_0$ the parabolic
confinement frequency.  Throughout this paper, we take the magnetic
field to be zero, and therefore set the vector potential
$\widehat{\vect{A}} \left( \hat{\vect{r}} \right) = 0$.

Two harmonic-oscillator quantum numbers, $n, m = 0, 1, 2, \ldots$,
characterize the eigenstates of the single-particle
Hamiltonian,\cite{jacak97:quant.dots} Eq.~(\ref{eq:nonIntHam}).  These
eigenstates are the ``atomic orbitals'' of the QD, and are given by
\begin{equation}
  \label{eq:eigenstate}
  \ket{n m} = \frac{1}{\sqrt{n! m!}}(a^\dagger)^n(b^\dagger)^m\ket{0 0},
\end{equation}
where, $a^\dagger$ and $b^\dagger$ are the usual Bose creation
operators, and $\ket{00}$ is the single-particle ground state.  These
orbitals have energy $\varepsilon_{nm}$ given by
\begin{equation}
  \label{eq:energy}
  \varepsilon_{nm} = \hbar \Omega_+ (n + \tfrac{1}{2}) + 
  \hbar \Omega_- (m + \tfrac{1}{2}),
\end{equation}
where $\Omega_\pm = (\sqrt{4 \omega_0^2 + \omega_c^2} \pm \omega_c) /
2$, and $\omega_c = e \vect{B} / (m^* c)$ is the cyclotron frequency.
This energy reduces to $\hbar\,\omega_0(n + m + 1)$ in the absence of
a magnetic field.

The single-particle Hamiltonian, the $z$-component of the orbital
angular momentum, $\hat{L}_z$, and a component of the spin
operator---which we take to be the $z$-component $\hat{S}_z$---form a
set of commuting observables which we take to classify our states:
$\hat{L}_z |nms\rangle = \hbar (n - m) |nms\rangle$, $\hat{S}_z
|nms\rangle = \hbar s |nms\rangle$, $\hat{h} |nms\rangle =
\varepsilon_{nm} |nms\rangle$.

We are interested in spatial textures formed by the spin field, and so
we require the position-space representation of the
orbitals.\cite{jacak97:quant.dots} These are given by\cite{orbitals}
\begin{multline}
  \label{eq:phiNM}
    \phi_{nm}(r, \theta) =  (-1)^{n_r} \frac{1}{\sqrt{2 \pi} l_0}
    \sqrt{\frac{n_r!}{\left(n_r + |m'|\right)!}}\\
    \times e^{i m' \theta} e^{-r^2 / (4 l_0^2)} 
    \left(\frac{r}{\sqrt{2} l_0} \right)^{|m'|} 
    L_{n_r}^{|m'|} \left(\frac{r^2}{2 l_0^2}\right),
\end{multline}
where $r$ and $\theta$ are the polar coordinates in two dimensions,
$l_0 = \sqrt{\hbar / (2 m^* \omega)}$ is the effective length with
$\omega = \sqrt{4 \omega_0^2 + \omega_c^2} / 2$, $n' = n + m$, $m' = n
- m$, $n_r=(n' - | m' |)/2$, and $L_n^{(\alpha)}(x)$ is the
generalized Laguerre polynomial.\cite{Abramowitz1965}

The eigenstates of the interacting system are determined by exact
diagonalization of Eq.~\eqref{eq:intHam}.  This procedure begins by
determining many-particle basis states (Slater determinants),
eigenstates of Eq.~\eqref{eq:nonIntHam}, that are composed of
antisymmetrized products of the single-particle states in
Eq.~\eqref{eq:eigenstate}.  We use 288 single-particle states, and as
many as 4900 many-particle basis states in the diagonalization
routine.  Without loss of fidelity, and for computational efficiency,
the number of many-particle basis states is reduced when determining
the two-point and three-point spin correlation calculations over the
range of the entire QD.  Block-diagonalization is performed for a
given set of parameters.  These include system parameters ($B$,
$\omega_0$, $m^*$, $\epsilon$) and the conserved quantities $N$,
$L_z$, $S_z$, $S^2$.  The Coulomb matrix elements are evaluated using
the convenient closed-form expression derived in
Ref.~\onlinecite{Kyriakidis02:Voltage-tunable-singlet-triplet}.

With the eigenstates determined, we calculate one-, two-, and
three-point position-dependent spin correlation functions over
energetically-degenerate manifolds.  The structure of the particular
operators used in these calculations is discussed in the next section.

\section{Product Spin Operators}
\label{sec:spinops}
For our investigation, we require the products of up to three one-body
spin operators.  Here, we introduce the product-operators used in the
calculations shown in the proceeding sections.  The details associated
with the derivation of each product-operator are discussed in the
Appendix.

\subsection{One-Body Spin Operators}
\label{sec:oneBody}
Except where indicated, all averages are taken over
energetically-degenerate manifolds.  That is to say,
\begin{subequations}
  \label{eq:averages}  
  \begin{equation}
    \label{eq:expectRho}
    \langle A \rangle = \frac{\text{Tr}
      (\hat{\rho}\,\hat{A})}{\text{Tr}(\hat{\rho})},
  \end{equation}
  where the density operator $\hat{\rho}$ is defined as
  \begin{equation}
    \label{eq:rho}
    \hat{\rho} = \sum_{i = 1}^n\ket{E_i}\bra{E_i},
  \end{equation}
\end{subequations}
and where the states $|E_i\rangle$ are all the states in a given
degenerate manifold of Eq.~(\ref{eq:hamil}): $\hat{{\cal H}}
|E_i\rangle = \text{const.} \, |E_i\rangle$ for all $i = 1, 2, \ldots
n$.

Our analysis begins with the evaluation of both spin density and
number density at position $\vect{r}$ in the QD system.  We are
specifically interested in isolating the spin-up and spin-down
densities along the coordinate axes.  Due to the conservation of spin
in the system, these one-body spin operators can only distinguish the
spin-up density from the spin-down density along a single axis.  We
therefore define a set of spin operators $\hat{S}_{\pm z}(\vect{r})$
that separately determines the spin-up and spin-down densities along
$z$ at position $\vect{r}$.  In canonical form (see Appendix
Eq.~\eqref{eq:oneBodyOp}), this is given by
\begin{equation}
  \label{eq:SPlusMinusZCanonical}
  \hat{S}_{s\,z}(\vect{r}) = \sum_{i\, j} U_{i j s}
  \hat{c}_{is}^\dagger \hat{c}_{js}^{\phantom{\dagger}},
\end{equation}
where $U_{i j s} = s \phi_i^*(\vect{r}) \phi_j(\vect{r}) / 2$, with $s
= \pm 1$ ($\hbar = 1$).  Note the composite indexes $i$ and $j$ each
represent a set of orbital quantum numbers $n$ and $m$.
 
The number density and spin density operators at position $\vect{r}$
are then given by
\begin{gather}
  \label{eq:n(r)}
  \hat{n}(\vect{r}) = \sum_s\hat{\psi}_s^\dagger(\vect{r})
  \hat{\psi}_s^{\phantom{\dagger}}(\vect{r}) = 
  2 \left( \hat{S}_{+z}(\vect{r}) + \hat{S}_{-z}(\vect{r}) \right),
  \\
  \label{eq:Sz}
  \hat{S}_z(\vect{r}) = \frac{1}{2}\sum_ss \hat{\psi}_s^\dagger(\vect{r})
  \hat{\psi}_s^{\phantom{\dagger}}(\vect{r}) = \left(
    \hat{S}_{+z}(\vect{r}) - \hat{S}_{-z}(\vect{r}) \right),
\end{gather}
where $\hat{\psi}_s^\dagger(\vect{r})$ and $\hat{\psi}_{s}(\vect{r})$
are field operators that respectively create and annihilate a fermion
at position $\vect{r}$, with spin $s$.  In terms of the eigenstates
$|nms\rangle$ of Eq.~(\ref{eq:nonIntHam}), the field operators are
given by
\begin{equation}
  \hat{\psi}_s^\dagger(\vect{r}) = \sum_{n\,m} \phi_{nm}^*(\vect{r})
  \hat{c}_{nms}^\dagger,
\end{equation}
with $\phi_{nm}(\vect{r}) = \langle \vect{r} | nm \rangle$ given in
Eq.~\eqref{eq:phiNM}.  

The effects of Coulomb interaction between the particles are apparent
when the expectation values of the above operators are compared
between the interacting and noninteracting systems.

\subsection{Two-Body Spin Operators}
\label{sec:twoBody}
We next investigate the two-point correlations as projected onto the
$z$-axis.  (Any other choice yields identical results.)  Unless
otherwise indicated, a spin-up (spin-down) particle refers to a
particle with spin polarized along the positive (negative) axis of
quantization (in this work, the $z$-axis).  Specifically, we
investigate $\langle S_{+z}(\vect{r}_0) S_{+z}(\vect{r}_1) \rangle$
and $\langle S_{+z}(\vect{r}_0) S_{-z}(\vect{r}_1) \rangle$.  In
canonical form, the operators are given by
\begin{subequations}
  \label{eq:twoPointOp}
  \begin{equation}
    \hat{S}_{+z}(\vect{r}_0) \hat{S}_{\sigma z}(\vect{r}_1) 
    = \frac{1}{2} \sum_{i j k l s s'} Q_{ijkl}^{s s' \sigma}
    \hat{c}_{is}^\dagger \hat{c}_{j s'}^\dagger 
    \hat{c}_{l s'} \hat{c}_{k s},
  \end{equation}
  with
  \begin{multline}
    Q_{ijkl}^{s s' \sigma} = \frac{1}{4} 
    \left(
      \phi_i^*(\vect{r}_0) \phi_j^*(\vect{r}_1) \phi_k(\vect{r}_0)
      \phi_l(\vect{r}_1) \delta_{s \uparrow} \delta_{s' \sigma} 
    \right.
    \\
    \left. \mbox{}
      + \phi_j^*(\vect{r}_0) \phi_i^*(\vect{r}_1) \phi_l(\vect{r}_0)
      \phi_k(\vect{r}_1) \delta_{s \sigma} \delta_{s' \uparrow}
    \right),
  \end{multline}
\end{subequations}
and $\sigma = \pm 1$.  As in Eq.~\eqref{eq:SPlusMinusZCanonical}, the
indexes $i$ through $l$ in Eq.~\eqref{eq:twoPointOp} are again
composite indexes of pairs of orbital quantum numbers.  The two-point
spin correlations measure the probability of finding a particle with
spin projection $\sigma z$ at position $\vect{r}_1$ given the
existence of a particle with spin projection $+z$ at position
$\vect{r}_0$.

Finally, the correlation between a spin-up particle at 
$\vect{r}_0$ and the \emph{net} spin density at $\vect{r}_1$ is
\begin{multline}
  \label{eq:twoPointOp_net}
  \hat{S}_{+z}(\vect{r}_0) \hat{S}_{\text{net } z}(\vect{r}_1) 
  \\ 
  = \hat{S}_{+z}(\vect{r}_0) \hat{S}_{+ z}(\vect{r}_1) 
  - \hat{S}_{+z}(\vect{r}_0) \hat{S}_{- z}(\vect{r}_1).
\end{multline}

These two-point spin correlations are useful for determining parallel
or antiparallel spin properties such as magnetic
ordering.\cite{Ghosal2007} They are insufficient, however, for
determining chiral textures where correlations are measured with
respect to orthogonal axes.  For that we turn to the three-point spin
correlation functions next.

\subsection{Three-Body Spin Operators}
\label{sec:threeBody}
We compute three unique three-point correlations:
\begin{subequations}
  \label{eq:3point}
  \begin{gather}
    \label{eq:xyx}
    \left\langle S_{+x}(\vect{r}_0) 
      S_{+y}(\vect{r}_1)
      S_{\text{net} \; x}(\vect{r}_2) \right\rangle,
    \\
    \label{eq:xyy}
    \left\langle S_{+x}(\vect{r}_0)
    S_{+ y}(\vect{r}_1)
    S_{\text{net} \; y}(\vect{r}_2) \right\rangle, 
    \\
\intertext{and}
    \label{eq:xyz}
    \left\langle S_{+x}(\vect{r}_0)
    S_{+ y}(\vect{r}_1)
    S_{\text{net} \; z}(\vect{r}_2) \right\rangle.
  \end{gather}
\end{subequations}
These three-point spin correlations measure the probability of finding
a particle with spin projected along the $x$, $y$ or $z$ axis,
respectively, given there is a particle at position $\vect{r}_0$ that
is spin-up along the $x$-axis \emph{and} a particle at position
$\vect{r}_1$ that is spin-up along the $y$-axis.  Whereas the
two-point functions can determine whether the spin-projection of a
second particle is parallel or antiparallel to the spin-projection of
the first particle, it cannot determine the orientation of the spin of
the second particle in a plane other than that of the spin of the
first particle.  The three-point functions in Eq. \eqref{eq:3point}
can indeed uncover such chiral structure.  Explicitly, the three-body
spin operators can be expressed as
\begin{subequations}
  \label{eq:3PtCanonical}
    \begin{multline} 
      \label{eq:3PtX_symm}
      \hat{S}_{+x}(\vect{r}_0)
      \hat{S}_{+y}(\vect{r}_1)
      \hat{S}_{\text{net}\;x}(\vect{r}_2) \\ = \frac{1}{3!} 
      \sum_{\substack{ijklmn\\s_{_1} s_{_2} s_{_3}}}
      K_{ijklmn}^{s_{_1} s_{_2} s_{_3}}
      \hat{c}_{is_{_1}}^\dagger \hat{c}_{js_{_2}}^\dagger 
      \hat{c}_{ks_{_3}}^\dagger  \hat{c}_{ns_{_3}}^{\phantom{\dagger}}
      \hat{c}_{ms_{_2}}^{\phantom{\dagger}}
      \hat{c}_{ls_{_1}}^{\phantom{\dagger}}, 
    \end{multline}
    where 
    \begin{multline}
      \label{eq:3PointCanonicalMatElem}
      K_{ijklmn}^{s_{_1} s_{_2} s_{_3}} = \frac{3}{16} 
      \delta_{s_{_1} s_{_2}} \delta_{s_{_2} \bar{s}_{_3}}
      \\ \times
      \left[ 
        i s_3 \phi_i^*(\vect{r}_0) \phi_j^*(\vect{r}_1) 
        \phi_k^*(\vect{r}_2) \phi_m(\vect{r}_0) 
        \phi_n(\vect{r}_1) \phi_l(\vect{r}_2) 
      \right.
      \\ \mbox{}
      - i \phi_i^*(\vect{r}_0) \phi_k^*(\vect{r}_1) 
      \phi_j^*(\vect{r}_2) \phi_m(\vect{r}_0) 
      \phi_l(\vect{r}_1) \phi_n(\vect{r}_2) 
      \\ \mbox{}
      -\phi_i^*(\vect{r}_0) \phi_j^*(\vect{r}_1) 
      \phi_k^*(\vect{r}_2) \phi_n(\vect{r}_0) 
      \phi_m(\vect{r}_1) \phi_l(\vect{r}_2)
      \\ 
      \left.
        \mbox{}
        - \phi_k^*(\vect{r}_0) \phi_i^*(\vect{r}_1) 
        \phi_j^*(\vect{r}_2) \phi_m(\vect{r}_0) 
        \phi_l(\vect{r}_1) \phi_n(\vect{r}_2) 
      \right],
    \end{multline}
\end{subequations}
and with similar expressions for the remaining two operators in
Eq.~(\ref{eq:3point}).  In Eq.~(\ref{eq:3PtCanonical}), each of the
six indexes $i$ through $n$ is once again a composite index over pairs
of orbital quantum numbers.  Furthermore, in
Eq.~(\ref{eq:3PointCanonicalMatElem}), $\bar{s}_i \equiv -s_i$, with
$s_i = \pm 1$ denoting the usual spin projections.

The choice for the first two spin-projections is not unique; due to
the absence of a preferred spin orientation in the system, each
expression is equivalent to any cyclic permutation of the spin
components.  We focus below on the cases where two of the three spins
operators lie in the ($x$-$y$) plane of the dot.

\section{Three-Particle System}
\label{N3}
In this section, we investigate spin correlations that exist in the
two lowest-lying states of a system with three charged particles.  Our
system is modeled with GaAs parameters ($\epsilon = 12.4$ and $m^* =
0.067m_e$), and our confinement potential is $\omega_0=1.0$~meV,
yielding an effective length at $B = 0$ of $l_0 = 23.8$~nm.

\subsection{Ground State Manifold}
\label{n3ground}
At zero magnetic field, the ground state of the three-particle system
is four-fold degenerate, with quantum numbers $L_z = \pm 1$, $S =
1/2$, and $S_z = \pm 1/2$.  We compute $\langle
S_{+z}(\vect{r})\rangle$ and $\langle S_{-z}(\vect{r})\rangle$ within
this degenerate subspace.  [See Eq.~\eqref{eq:averages}.]  From these,
we determine both the net density, $\langle n(\vect{r})\rangle$,
Eq.~(\ref{eq:n(r)}), and the net spin $\langle S_z (\vect{r})
\rangle$, Eq.~(\ref{eq:Sz}).  We then go on to calculate the two-point
spin functions to demonstrate correlations between parallel and
antiparallel spin components, followed by the three-point functions to
uncover chiral correlations.

\subsubsection{Single-particle densities}
\label{sec:single-part-dens}

To illustrate the effects of long-range Coulomb repulsions, we
consider the spin density with and without interactions.  In the
non-interacting case, each eigenstate is a single antisymmetrized
orbital configuration.  For the three-particle system, there are two
particles on the $|n m\rangle = |00\rangle$ orbital, and one on either
the $|10\rangle$ or $|01\rangle$ orbital.  This yields four degenerate
states with quantum numbers $(S, S_z, L_z) = (1/2, \pm 1/2, \pm 1)$.
For the interacting case, these symmetries are not explicitly broken;
the degeneracy and the quantum numbers remain the same, but the states
themselves are now correlated, involving many other orbital
configurations consistent with the symmetry.

\begin{figure}
  \centering
  \resizebox{\fig}{!}{\includegraphics{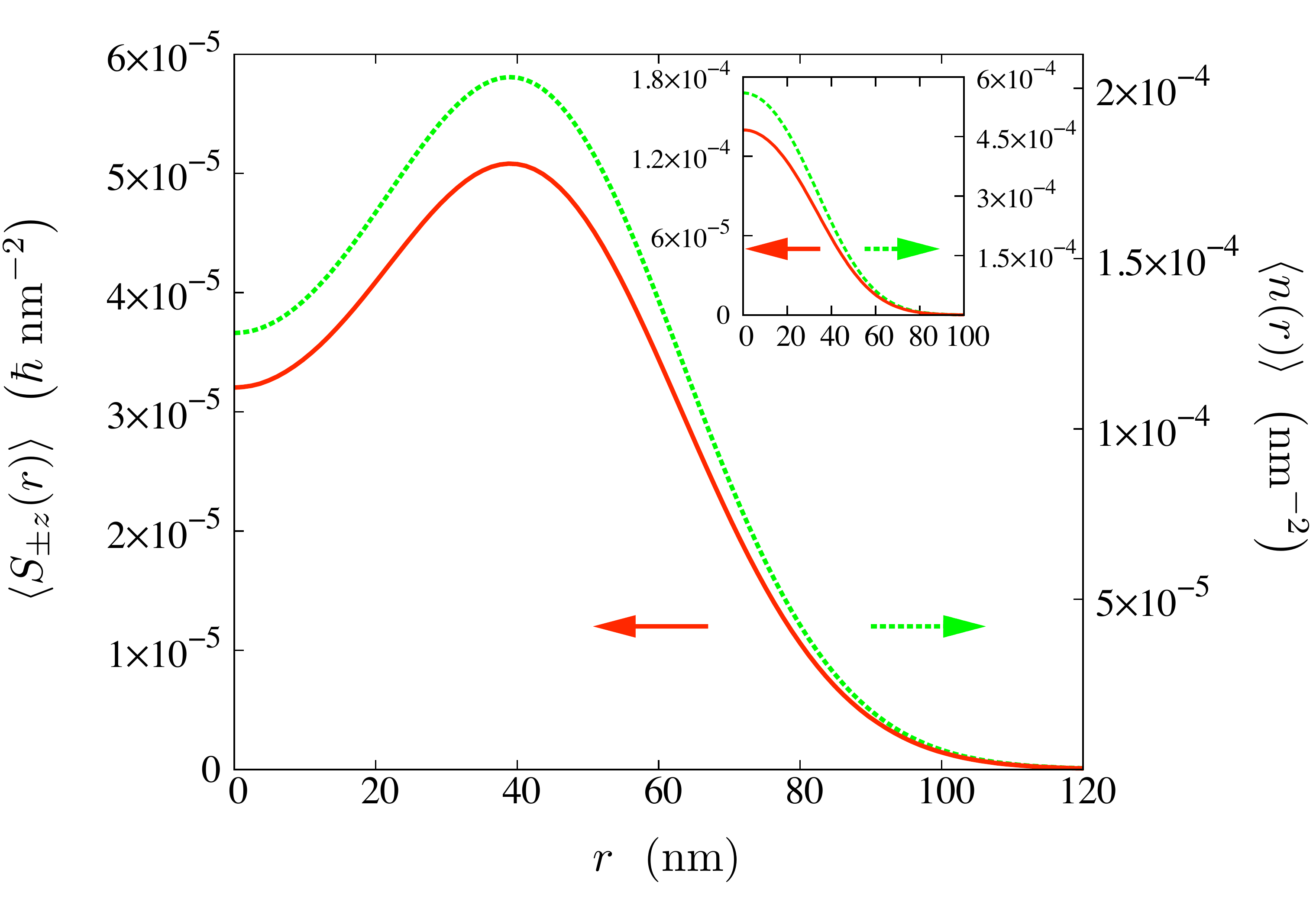}}

  \caption{\label{fig:1PtN3GS} (Color online.) Radial dependence of
    the single-particle spin and number density in the three-particle
    ground state manifold.  Inset shows densities in absence of
    Coulomb interaction.  (Inset axes have the same dimensions as the
    main plot).}
\end{figure}

Figure~\ref{fig:1PtN3GS} shows single-particle densities for the total
ground-state manifold as a function of radial distance from the center
for both the interacting and non-interacting cases.
There is azimuthal symmetry for these configurations due to the
underlying circular symmetry of the dot itself, manifest in the
Hamiltonian, Eq.~(\ref{eq:hamil}).

The non-interacting case is characterized as Gaussian-like with a peak
at the origin.  When Coulomb interactions are considered, the
repulsion smears out the density over different orbital
configurations.\cite{Stevenson11:Fractional-orbital} The competition
between repulsion and confinement results in an annular density about
the origin.  These interaction effects are strong; the ground-state
energy of the interacting system is 10.30~meV for these
experimentally-relevant system parameters---more than twice the
ground-state energy of the non-interacting case (4.0 meV).

The effects of Coulomb repulsion are also reflected in the
single-particle spin densities as well.  Note, however, that in both
cases we have $\langle S_{+z}(\vect{r}) \rangle = \langle
S_{-z}(\vect{r}) \rangle$: $\langle S_{\text{net }z} (\vect{r})
\rangle = 0$ everywhere in the dot, therefore the one-point
calculations are insufficient for showing the Coulomb effects on spin.
This is a consequence of the SU(2) symmetry present at zero magnetic
field.

\subsubsection{Two-point spin correlations}
\label{sec:two-point-spin}

For the two-point spin correlations, Eq.~\eqref{eq:twoPointOp}, we
consider the case where $\vect{r}_0$ is fixed at the location of
maximum single-particle density, $r_{\text{max}} \equiv 39$~nm, as
obtained in the previous section.  We further define the angular
location of $\vect{r}_0$ to be $\theta_0 = 0$.  We do not the discuss
the non-interacting limit for these calculations.

\begin{figure}
  \centering
  \resizebox{\fig}{!}{\includegraphics{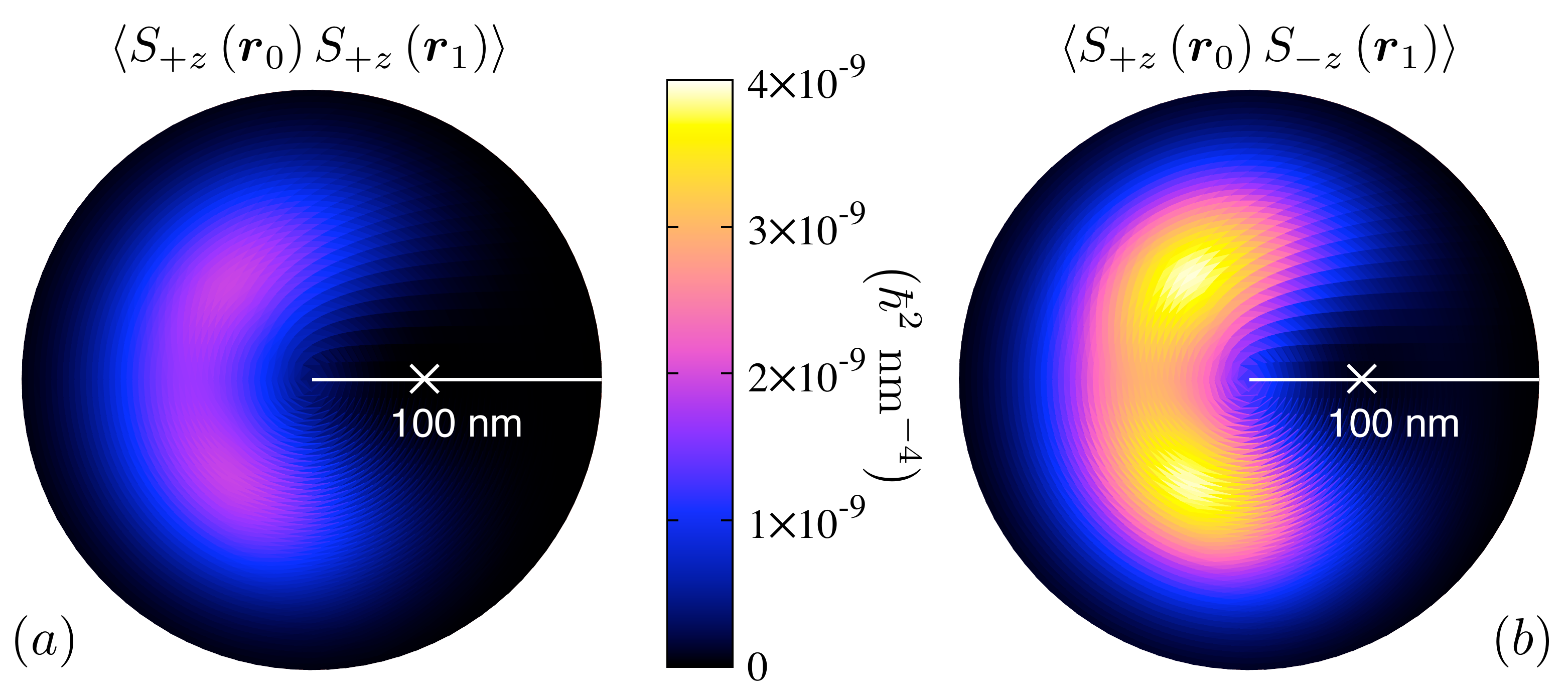}}
  \caption{\label{fig:2PtN3GS} (Color online.)  Two-point spin
    correlation functions for the three-particle ground state
    manifold.  Distributions are shown for the spin-up (a) and
    spin-down (b) densities at $\vect{r}_1$ given a spin-up particle
    at $\vect{r}_0 = (39~\text{nm},\, 0)$, denoted by a cross.}
\end{figure}

The two-point correlations $\langle S_{+z}(\vect{r}_0)
S_{+z}(\vect{r}_1)\rangle$ and $\langle S_{+z}(\vect{r}_0)
S_{-z}(\vect{r}_1)\rangle$ for the interacting ground-state manifold
are shown in Fig.~\ref{fig:2PtN3GS} as a function of $\vect{r}_1$ for
$\vect{r}_0 = (r_{\text{max}}, \theta_0)$.
Two peaks are evident along the ring of radius $r_{\text{max}}$.  Note
that our averages, Eq.~(\ref{eq:averages}), are obtained by tracing
over \emph{all} the degenerate states in the ground-state manifold.
An incipient Wigner crystallization is apparent with the spins forming
a classical-like lattice at the vertexes of a
triangle.\cite{Ghosal2006,Ghosal2007} This structure is not seen in
the non-interacting case, implying that the Coulomb repulsion between
the particles is responsible for this spin texture.

\begin{figure}
  \centering
  \resizebox{\smallfig}{!}{\includegraphics{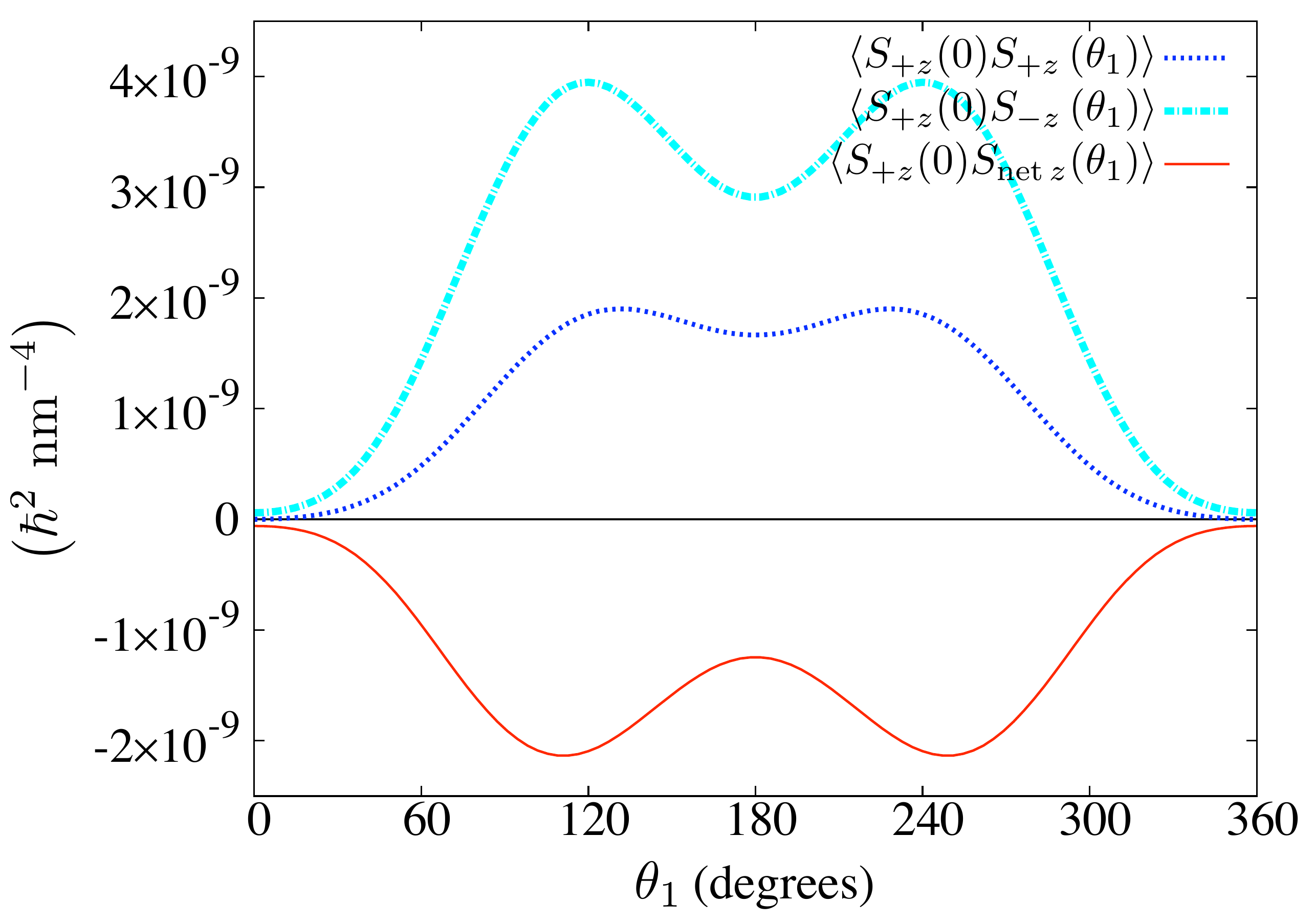}}
  \caption{\label{fig:2PtN3GS_net} (Color online.) Trace along
    $r_{\text{max}} = 39$~nm in the three-particle ground state
    manifold, revealing the spin distribution with respect to a
    spin-up particle at $\vect{r}_0 = (r_{\text{max}}, 0)$.}
\end{figure}

To more clearly probe the angular inhomogeneity, we plot in
Fig.~\ref{fig:2PtN3GS_net} results along the ring $r_{\text{max}}$.
In particular, we show the net spin $S_z$ as well as the individual
components $S_{\pm z}$, given a spin-up particle at $\vect{r}_0$.
Note as $\vect{r}_1$ approaches $\vect{r}_0$ the remaining spin-up
density goes to zero, indicative of a Pauli
vortex\cite{Saarikoski2010} at that position. As well, the spin
density at the two peaks is not fully polarized, indicating a degree
of canting away from the $z$-axis: The net spin tilts towards the
$x$-$y$ plane.  The lack of equal magnitudes of spin-up and spin-down
probabilities at every point along $r_{\text{max}}$ in
Fig.~\ref{fig:2PtN3GS_net} indicates that the spin density never lies
completely in the $x$-$y$ plane.  Since the spin-density never crosses
through the plane, it cannot have winding order.  Windings are
important in these spin systems as they represent clear examples of
topologically stable structures.

\begin{figure}
  \centering
  \resizebox{\smallfig}{!}{\includegraphics{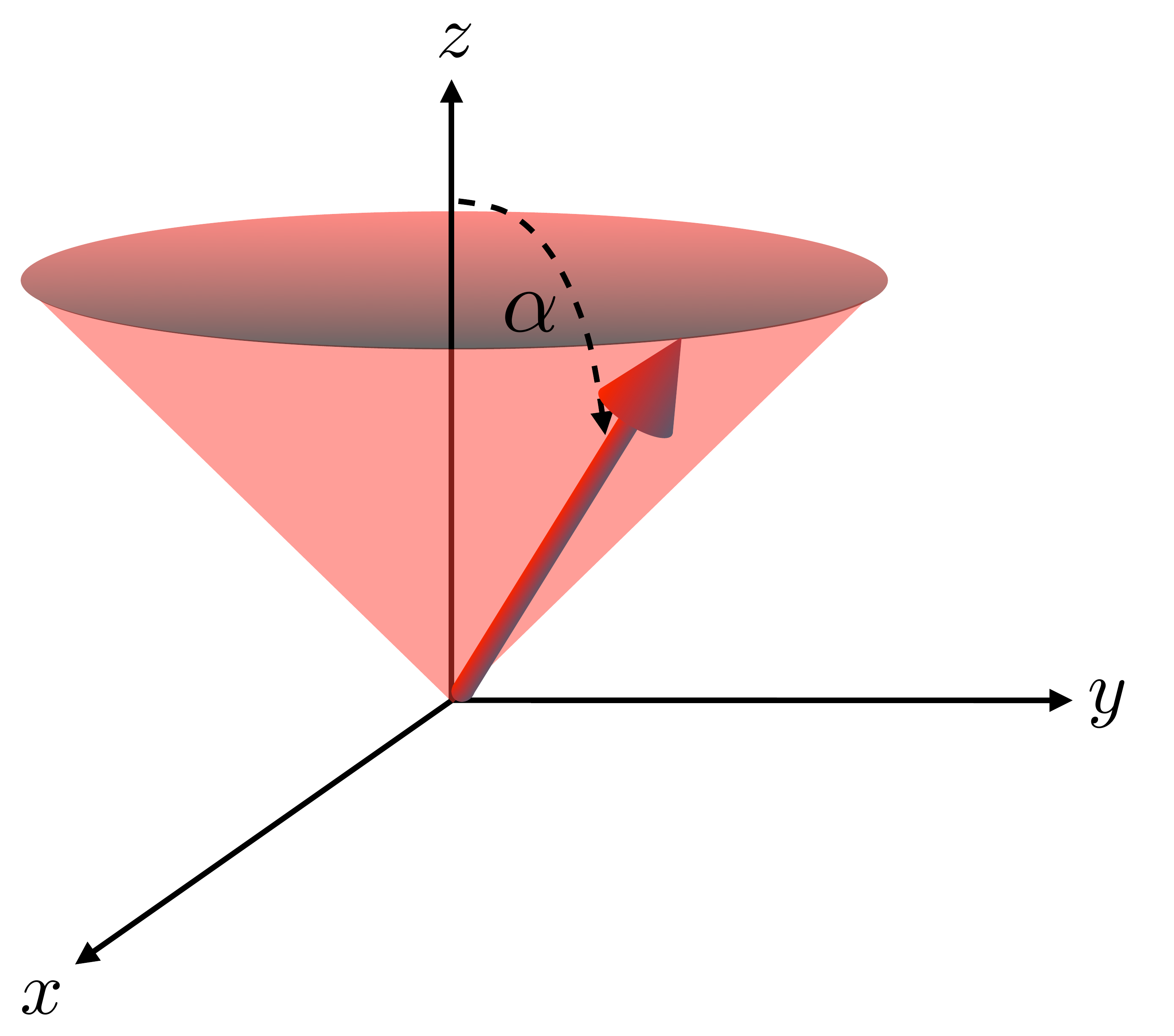}}
  \caption{\label{fig:canting} (Color online.) Canting angle $\alpha$
    with respect to the axis of quantization (the $z$ axis).  The spin
    is not fully polarized along either the $+z$ or $-z$ axis but has
    components of both projections, thus smearing out along a cone of
    probability about the z-axis.}
\end{figure}

The two-point correlations are insufficient to determine the probable
orientation of the spin in the x-y plane, so the results of these
calculations can be interpreted as the smearing out of the net spin
across the surface of a cone centred at a local z-axis at every point
measured in the calculation.  The opening angle of the cone is twice
that of the local canting angle, the degree of tilting from the local
z-axis, of the spin.  This cone, along with the canting angle, is
illustrated in Fig.~\ref{fig:canting}.

The canting angle can be determined in the following manner: If we
consider the spin density field shown in Fig.~\ref{fig:2PtN3GS_net} as
itself a spin-half field, we may generally write its local orientation
as
\begin{equation}
  \label{eq:bloch-vect}
  \left| \psi(r_{\text{max}}, \theta_1) \right\rangle = \frac{1}{N}
  \left(c_+ |+z\rangle + \text{e}^{i \beta} c_- |-z\rangle\right),
\end{equation}
where $c_\pm$ are real and may be defined as
\begin{subequations}
  \label{eq:bloch-consts}
  \begin{gather}
    c_+ = 
    \left\langle 
      S_{+z}(\vect{r}_0) S_{+z}(r_{\text{max}}, \theta_1)
    \right\rangle
    \equiv N \cos \frac{\alpha}{2}, \\
    c_- = 
    \left\langle 
      S_{+z}(\vect{r}_0) S_{-z}(r_{\text{max}}, \theta_1)
    \right\rangle
    \equiv N \sin \frac{\alpha}{2},
  \end{gather}  
\end{subequations}
and where $N = \sqrt{c_+^2 + c_-^2}$ is a local normalization.  The
symmetry of the two-point functions prevents discrimination of
different values of the azimuthal angle $\beta$, but it \emph{can}
determine the canting angle $\alpha$.

\begin{figure}
  \centering
  \resizebox{\smallfig}{!}{\includegraphics{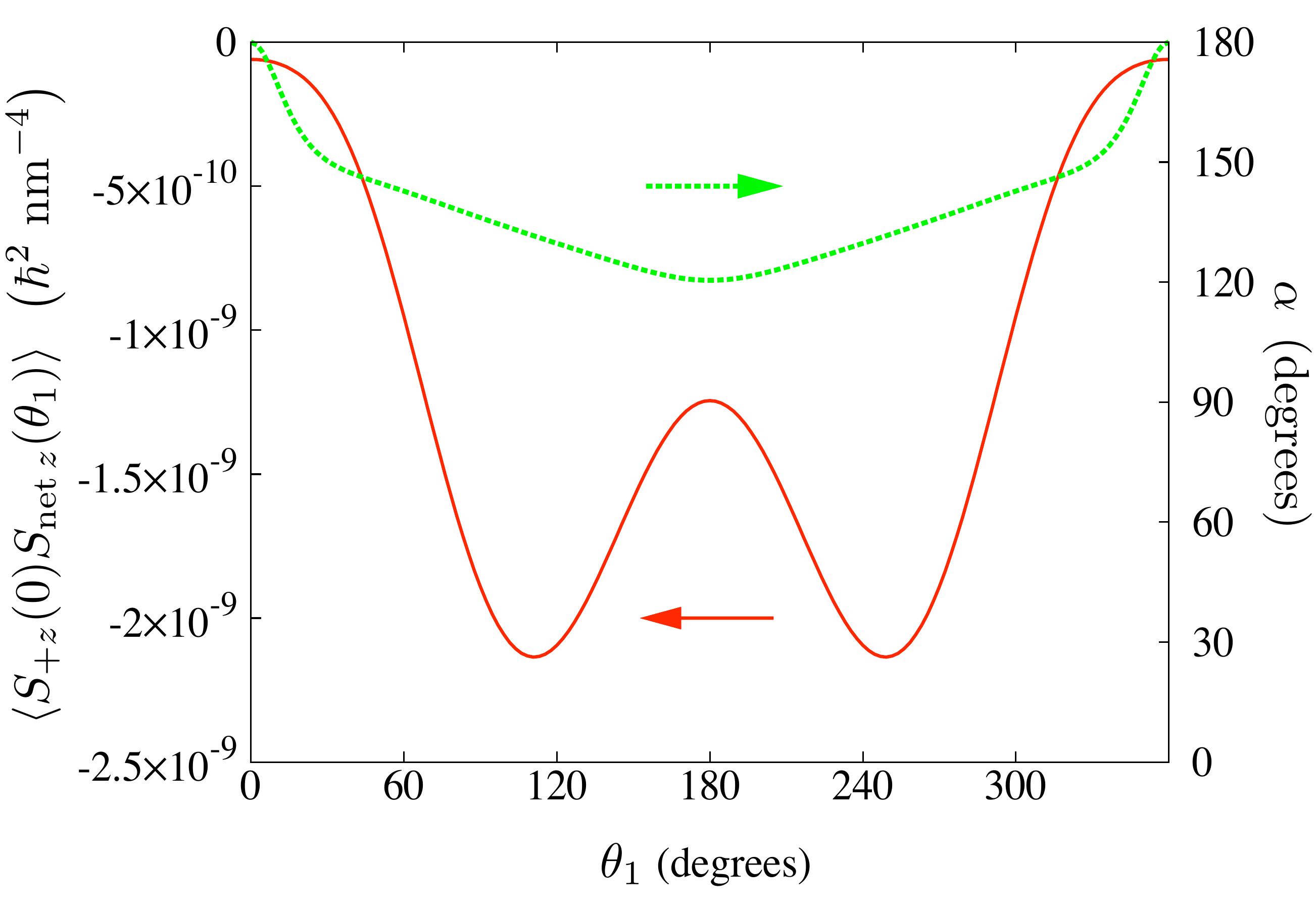}}
  \caption{\label{fig:N3GScanting} (Color online.) Canting angle
    $\alpha$ as a function of $\theta_1$ along the ring of maximum
    single-particle density $r_{\text{max}}$ = 39 nm in the
    three-particle ground state manifold.  The net spin density from
    the two-point calculation along the same ring $r_{\text{max}}$ is
    shown for reference.}
\end{figure}

For each of the two peaks in Fig.~\ref{fig:2PtN3GS_net}, the canting
angle with respect to the positive $z$-axis is determined to be
$\alpha = 131^\circ$.  That is, relative to a spin-up particle at
$\vect{r}_0$, the spin-density peaks describing the other two
particles both occupy the surface of a cone with canting angle
49$^\circ$ from the negative $z$-axis.  Figure \ref{fig:N3GScanting}
shows the resulting canting angle from the positive $z$-axis of the
local Bloch vector for each value of $\theta_1$ along the ring
$r_{\text{max}}$.  The canting angle becomes greatest ($\alpha$
approaches $180^\circ$) as $\theta_1$ approaches $\theta_0$ = 0, and
is a minimum at $\theta_1 = 180^\circ$.  (Pauli exclusion dictates
that $\alpha \rightarrow 180^\circ$ as $\theta_1 \rightarrow
\theta_0$.)

The two-point functions in this system with SU(2) symmetry are
insufficient to distinguish chiral structures.  Three-point functions
are necessary to resolve spin components in the plane perpendicular to
the axis defined by the two-point functions.  We turn to these next.

\subsubsection{Three-point spin correlations}
\label{sec:three-point-spin}

As described in Sec.~\ref{sec:threeBody}, we compute the three
distinct three-point correlation functions given in
Eq.~(\ref{eq:3point}).  Other three-point functions can be related to
one of these three due to the symmetry of the underlying Hamiltonian.

\begin{figure}
  \centering
  \resizebox{\fig}{!}{\includegraphics{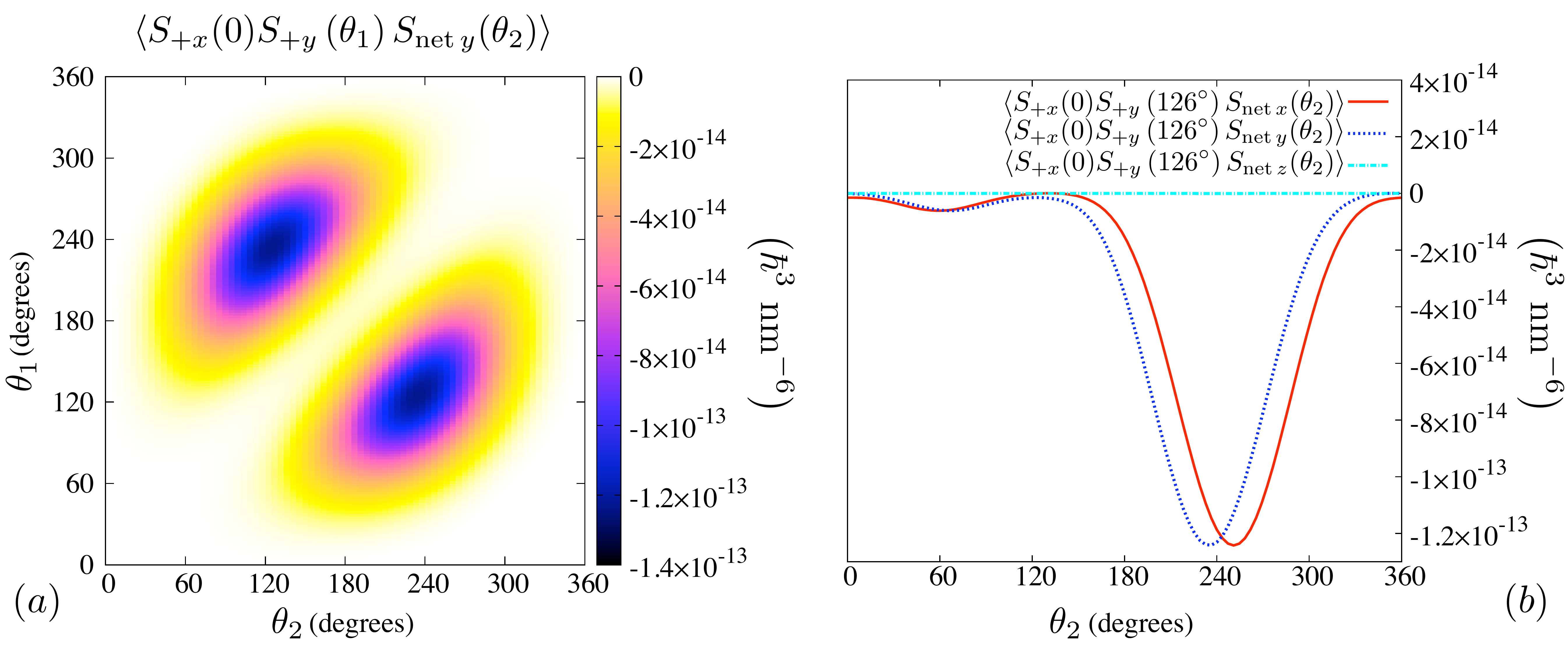}}
  \caption{\label{fig:3PtN3GS_square} (Color online.) Three-point spin
    correlations, Eq.~(\ref{eq:3point}), along the ring
    $r_{\text{max}} = 39$~nm for the three-particle ground state
    manifold.  (a) The net $S_y$ is plotted given a spin-up particle
    along the $x$-axis at $\theta_0=0^\circ$ and a spin-up particle
    along the $y$-axis at $\theta_1$.  (b) The three-point
    correlations for net spins along the $x$, $y$ and $z$ axes for
    $\theta_0 = 0^\circ$ and $\theta_1 = 126^\circ$ on the ring
    $r_{\text{max}}$.}
\end{figure}

In Eq.~(\ref{eq:3point}), we fix $\vect{r}_0$ on the ring of maximum
single-particle density ($r_{\text{max}}$ = 39 nm) at $\theta_0$ = 0.
The correlation functions are then a function of the four variables
$r_1$, $\theta_1$, $r_2$, and $\theta_2$.  If we further choose to
probe the system along the ring $r_{\text{max}}$, we then obtain the
two-dimensional map in the two angles $\theta_1$ and $\theta_2$ shown
in Fig.~\ref{fig:3PtN3GS_square}(a).
Explicitly, Fig.~\ref{fig:3PtN3GS_square}(a) is a plot of $\langle
S_{+x}(r_{\text{max}},\, 0) S_{+y}(r_{\text{max}}, \theta_1)
S_{\text{net}\ y}(r_{\text{max}}, \theta_2)\rangle$ as a function of
$\theta_1$ and $\theta_2$.  As the angular location of the second spin
approaches $\theta_1 = 120^\circ$ with respect to the first spin, a
peak emerges in the net spin distribution along the $y$-axis.  The
maximum correlations occur at $(\theta_1, \theta_2) = (126^\circ,
235^\circ)$ and $(234^\circ, 125^\circ)$.  Note the inversion symmetry
about the point $(180^\circ, 180^\circ)$.  Very similar results are
seen for the net spin distribution along the $x$-axis (not shown).

The three-point correlations for the net spin distributions along the
$x$, $y$ and $z$ axes with respect to a spin-up particle along the
$x$-axis at $\vect{r}_0 = (r_{\text{max}}, 0)$, and a spin-up particle
along the $y$-axis at $\vect{r}_1 = (r_{\text{max}}, 126^\circ)$ are
displayed in Fig.~\ref{fig:3PtN3GS_square}(b). Note the net spin
distribution along the $z$-axis is negligible in comparison to the
distributions along the $x$ and $y$ axes: Any spin density that
remains in the system lies only in the plane of the two spins at
$\vect{r}_0$ and $\vect{r}_1$, respectively.

\begin{figure}
  \centering
  \resizebox{\smallfig}{!}{\includegraphics{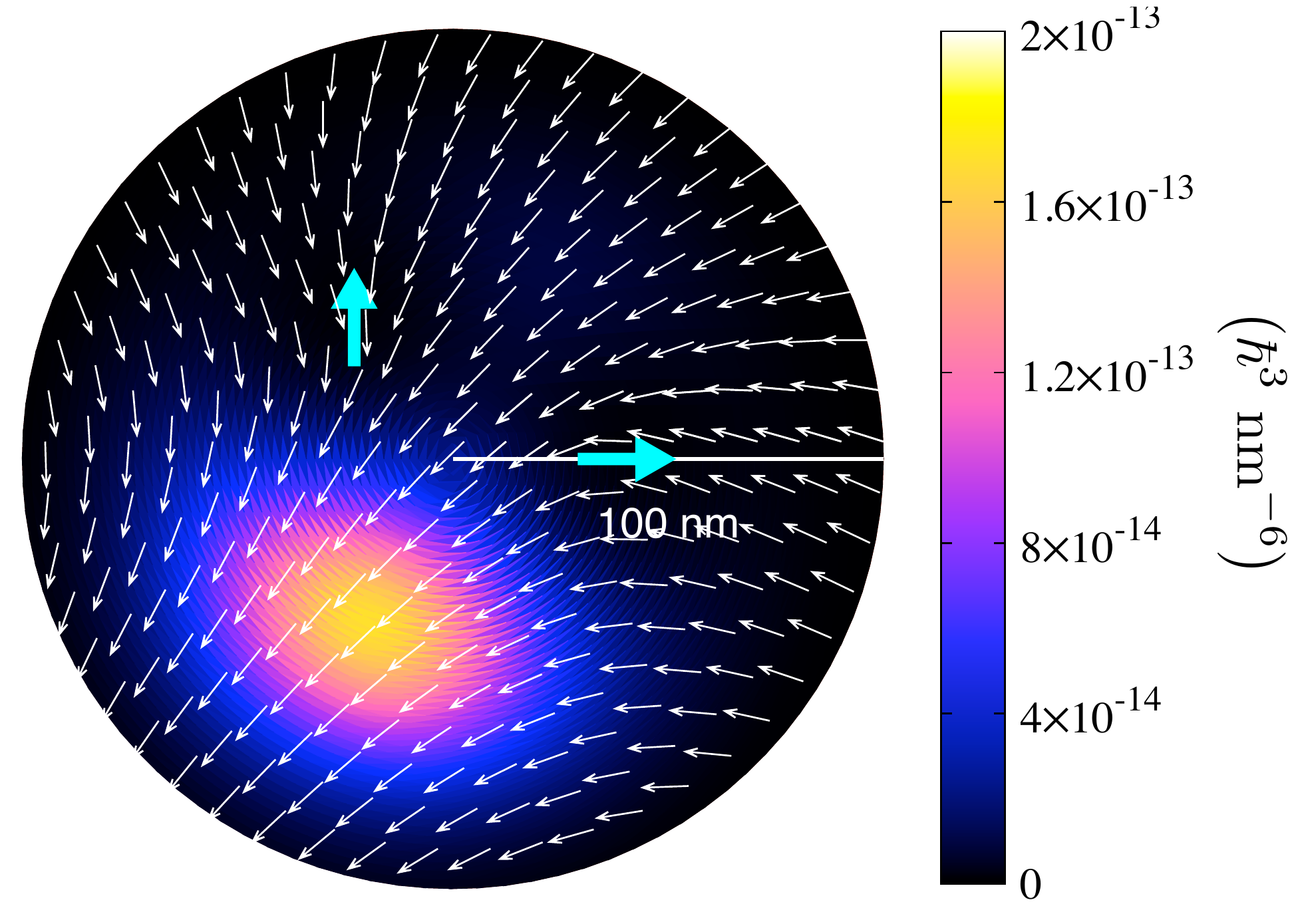}}
  \caption{\label{fig:3PtN3GS_map}(Color online.)  Net spin
    distribution in the plane of the QD for the three-particle ground
    state manifold, as determined by three-point spin correlations.
    The magnitude is denoted by the color bar and the direction is
    indicated by the vector field.  The $z$-component is negligible
    here.  The calculation is done with respect to a particle of
    spin-up projection on the $x$-axis at $\vect{r}_0=(39~\text{nm},
    0^\circ)$ and a particle of spin-up projection on the $y$-axis at
    $\vect{r}_1=(39~\text{nm}, 126^\circ)$.  (These locations are
    denoted by the blue arrows.}
\end{figure}

Figure~\ref{fig:3PtN3GS_map} is a map of the net spin distribution in
the three-particle ground-state manifold, given a spin-up particle
along $x$ at $\vect{r}_0 = (r_{\text{max}},\, 0)$ and a spin-up
particle along $y$ at $\vect{r}_1 = (r_{\text{max}}, 126^\circ)$.
The peak along the ring $r_{\text{max}}$ occurs at $\theta_3 =
243^\circ$, and the net spin at this peak has an equal spin-down
projection along each the $x$ and $y$ axes.  Note in this distribution
the peak of the net spin density and the two locations $\vect{r}_0$
and $\vect{r}_1$ are each equidistant from each other.

The three-point correlations suggest that the most probable spin
configuration in this three-particle state has a planar, splayed
order, as seen in Fig.~\ref{fig:3PtN3GS_map}.  Recall, however, that
the results from the two-point correlations imply that the state does
not have a clear winding order.

\subsection{First Excited State}
\label{n3one}
Like the three-particle ground state, the first excited state is also
four-fold degenerate.  However, in contrast to the ground state
manifold, $L_z = 0$ for all four degenerate states in the first
excited state.  Degeneracy occurs through the spin quantum numbers $S
= 3/2$, and $S_z = \pm 3/2,\ \pm 1/2$.

As in the ground state, the single-particle density and the spin
density distributions are rotationally symmetric.  When interactions
are considered, the annular distributions peak at $r_{\text{max}}
\equiv 41$~nm.  These distributions are similar to those seen in
Fig.~\ref{fig:1PtN3GS} for the ground state and are not shown.  As
seen in the ground state manifold, the net $S_z$ is once again zero
everywhere in the QD for both the interacting and non-interacting
cases.  The non-interacting limit is not investigated further.

The two-point spin correlations in the interacting first excited state
are similar to those seen for the interacting ground state (see
Fig.~\ref{fig:2PtN3GS}), namely, there are two peaks along the ring
$r_{\text{max}}$.  The two-point distribution also shows canting
towards the x-y plane with respect to a spin-up particle at
$\vect{r}_0 \equiv (r_{\text{max}},\, 0)$.  Unlike the ground state,
this canting is constant for all $\theta$ along the ring
$r_{\text{max}}$ except at a very small region about $\theta = 0$ in
accordance with Pauli exclusion, and has a value of $\alpha =
53^\circ$.  Once again a Pauli vortex is present at $\vect{r}_0$.

We turn now briefly to the three-point spin correlations in the first
excited state along the ring $r_{\text{max}}$.  We fix one spin-up
particle along the $x$-axis at $\theta_0 = 0^\circ$, and the other
spin-up particle along the $y$-axis at $\theta_1 = 120^\circ$, the
location of one of the two peaks determined by the two-point
correlation calculation.  Due to the spin polarized states, minor
differences exist between this manifold and the ground state manifold,
but otherwise these results are very similar and so results for the
excited state manifold are not shown.

\section{Four-Particle System}
\label{N4}

We examine the lowest two energy eigenstates in a system of four
charged particles for the same system parameters used above.  We begin
the investigation of the four-particle system by first determining the
radial location of maximum single-particle density in each state.
With this established, we calculate higher-order spin correlations in
this region.

\subsection{Ground State}
\label{n4ground}

The ground state of the four-particle system is three-fold degenerate
with quantum numbers $L_z = 0$, $S = 1$, and $S_z = \pm 1,~0$.  The
single-particle density distribution in the ground state manifold (not
shown) is annular in shape with a peak at $r_{\text{max}} \equiv
48$~nm from the center of the dot.  This distance is greater than in
the three-particle states primarily due to the additional Coulomb
repulsion present in the system.

\begin{figure}
  \centering
  \resizebox{\fig}{!}{\includegraphics{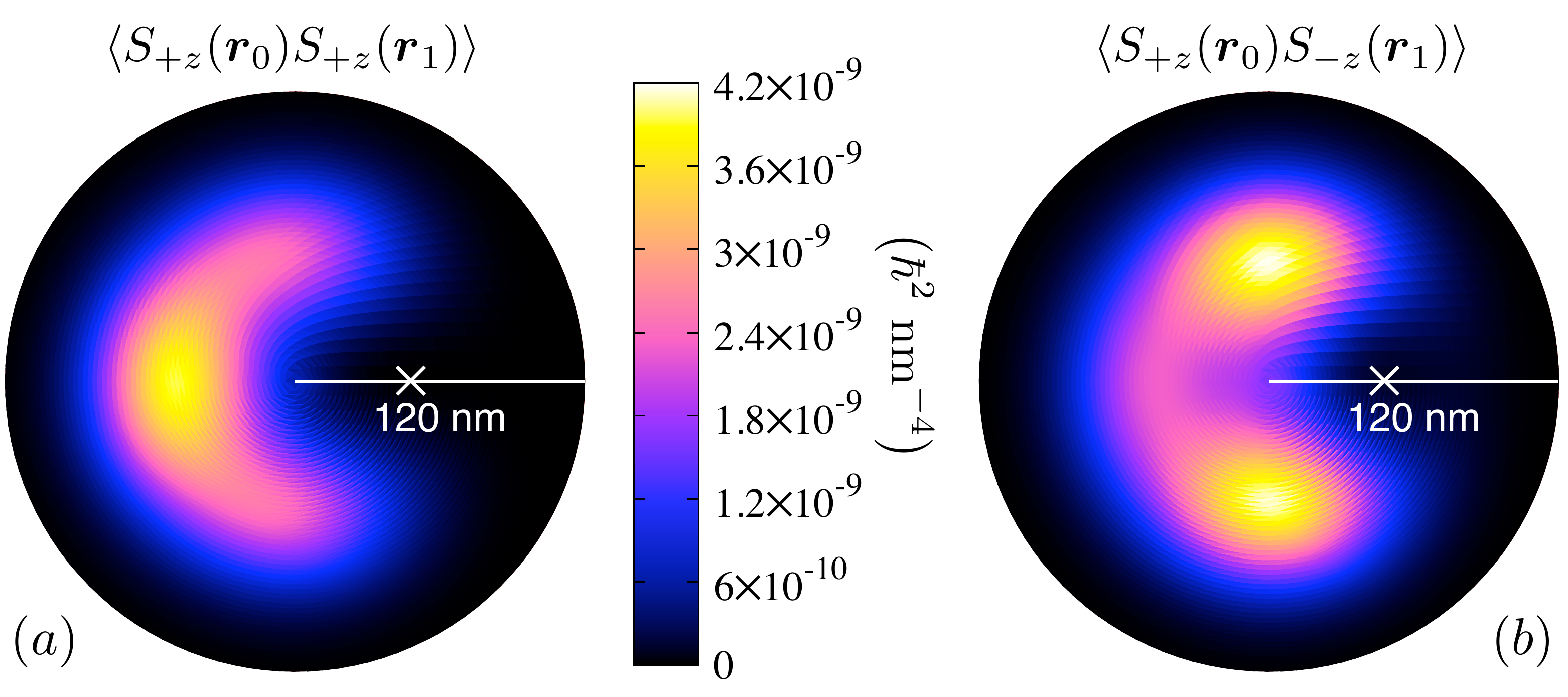}}
  \caption{\label{fig:N42PtGS}(Color online.)  Two-point spin
    correlations in the four-particle ground state manifold given a
    spin-up particle at $\vect{r}_0 = (48 \text{~nm},\, 0)$ (white
    cross). The distributions are shown for the net spin-up density
    (a) and spin-down density (b).}
\end{figure}

\begin{figure}
  \centering
  \resizebox{\fig}{!}{\includegraphics{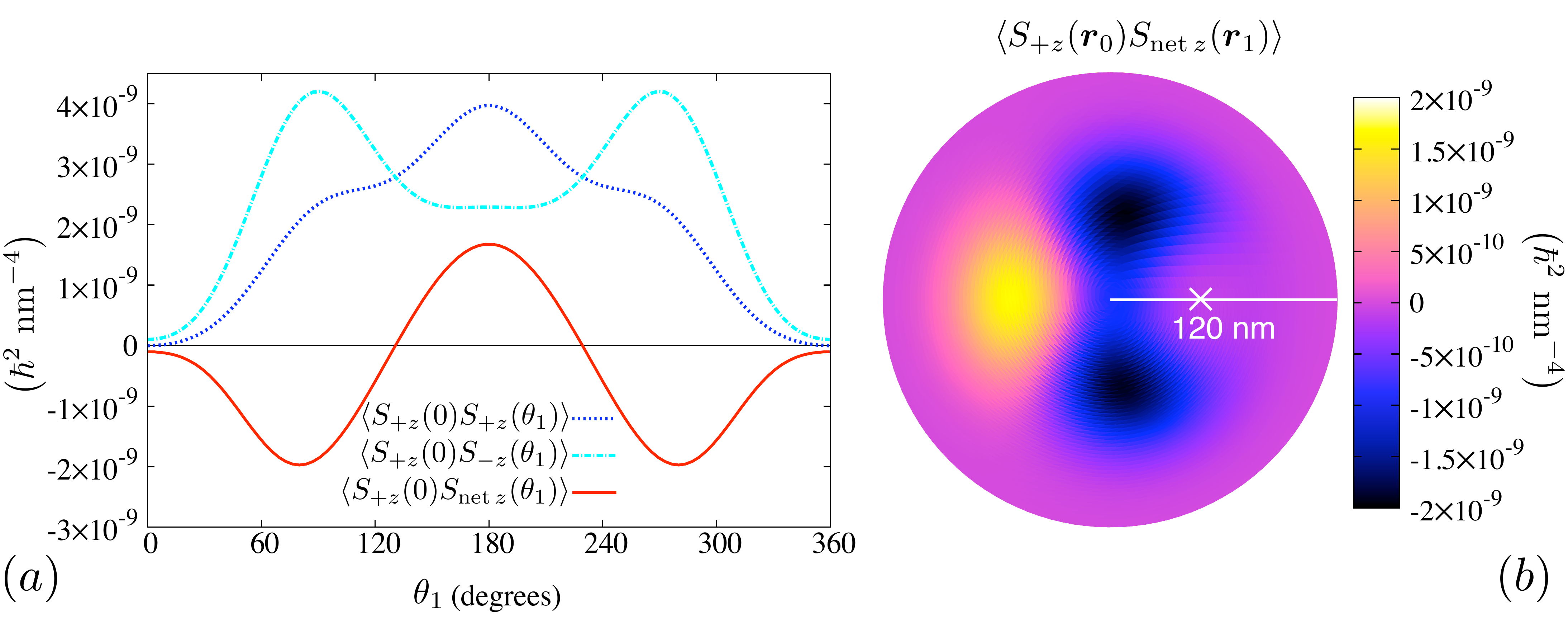}}
  \caption{\label{fig:N42PtGS_net} (Color online.)  Two-point spin
    correlations in the four-particle ground state manifold.  (a)
    Trace is along the ring of maximum single-particle density
    $r_{\text{max}}$ = 48 nm, with respect to a spin-up particle
    located at $\vect{r}_0=(r_{\text{max}}, 0^\circ)$.  (b) Net spin
    in the QD, given a spin-up particle at
    $\vect{r}_0=(r_{\text{max}}, 0^\circ)$.}
\end{figure}

The two-point spin correlations in the four-particle ground state are
calculated with respect to a spin-up particle at $\vect{r}_0 \equiv
(r_{\text{max}},\, 0)$.  The calculations are similar to those for the
three-particle system.  Figure~\ref{fig:N42PtGS} shows the
distribution of the remaining spin-up density and spin-down density in
the ground state manifold with respect to the spin-up particle at
$\vect{r}_0$.
As seen in the three-particle ground state, there is a Pauli vortex at
$\vect{r}_0$.  Figure~\ref{fig:N42PtGS}(a) shows that the probability
of finding another spin-up particle is strongest at $\theta_1 =
180^\circ$.  Conversely, the spin-down distribution in
Fig.~\ref{fig:N42PtGS}(b) shows two peaks, one at $\theta_1 =
90^\circ$ and the other at $\theta_1 = 270^\circ$.  The saddle point
at $\theta_1=180^\circ$ is more than half of the magnitude of the
peaks. Taken together, the two plots in Fig.~\ref{fig:N42PtGS}
indicate an antiferromagnetic alignment of the spins, with each spin
equidistant from each other, distributed along the ring of maximum
single-particle density.  This is further revealed in the net spin
distribution, shown in Fig.~\ref{fig:N42PtGS_net}.
Figure~\ref{fig:N42PtGS_net}(b) shows the evident antiferromagnetic
tendency in the four-particle ground-state manifold.  However, as
shown in Fig.~\ref{fig:N42PtGS_net}(a), in this small system, the
spins are neither fully localized nor fully polarized in the quantum
dot.

\begin{figure}
  \centering
  \resizebox{\smallfig}{!}{\includegraphics{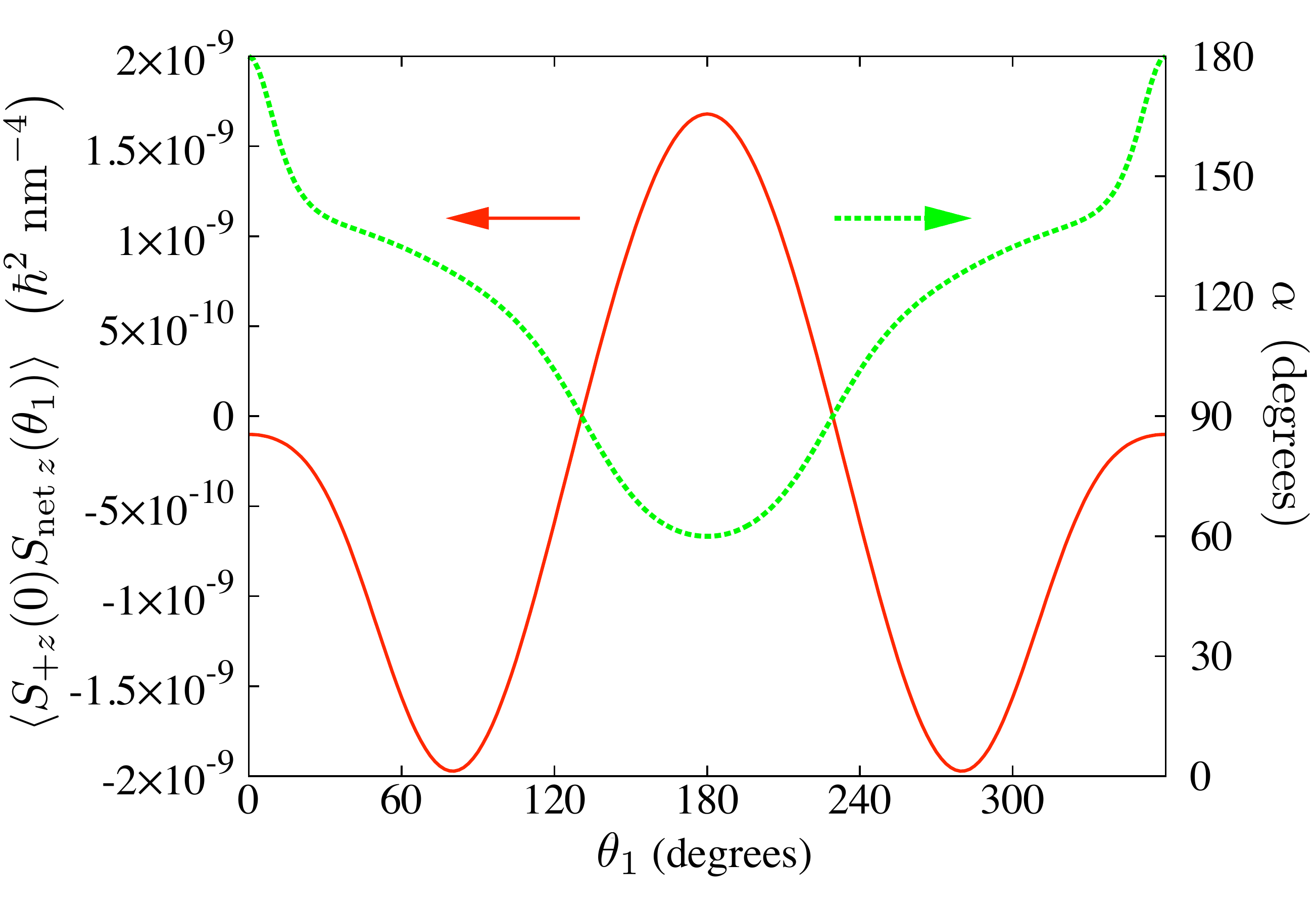}}
  \caption{\label{fig:N4GScanting} (Color online.) Canting angle
    $\alpha$ as a function of $\theta_1$ along the ring of maximum
    single-particle density $r_{\text{max}}$ = 48 nm in the
    four-particle ground state manifold.  The net spin density from
    the two-point calculation along the same ring $r_{\text{max}}$ is
    shown for reference.}
\end{figure}

In the four-particle ground-state manifold, the net spin density from
the two-point calculation shows canting along the ring
$r_{\text{max}}$ as a function of $\theta_1$ similar to that in the
three-particle ground-state manifold.  The canting angle becomes
antiparallel ($\alpha = 180^\circ$) relative to the spin-up particle
at $\vect{r}_0 = (r_{\text{max}},\, 0)$ as $\theta_1$ approaches zero,
passing through the $x$-$y$ plane when the net spin density from the
two-point function is zero along the ring $r_{\text{max}}$ (near
$\theta_1 = 130^\circ$ and $\theta_1 = 230^\circ$), and has a positive
projection on the $z$-axis at $\theta_1 = 180^\circ$, where the
two-point function shows the net spin density to be predominantly
spin-up.  The canting angle along $r_{\text{max}}$ is consistent with
the antiferromagnetic ordering suggested by the two-point correlation
calculations.  The result is shown in Fig.~\ref{fig:N4GScanting}.

\begin{figure}
  \centering
  \resizebox{\fig}{!}{\includegraphics{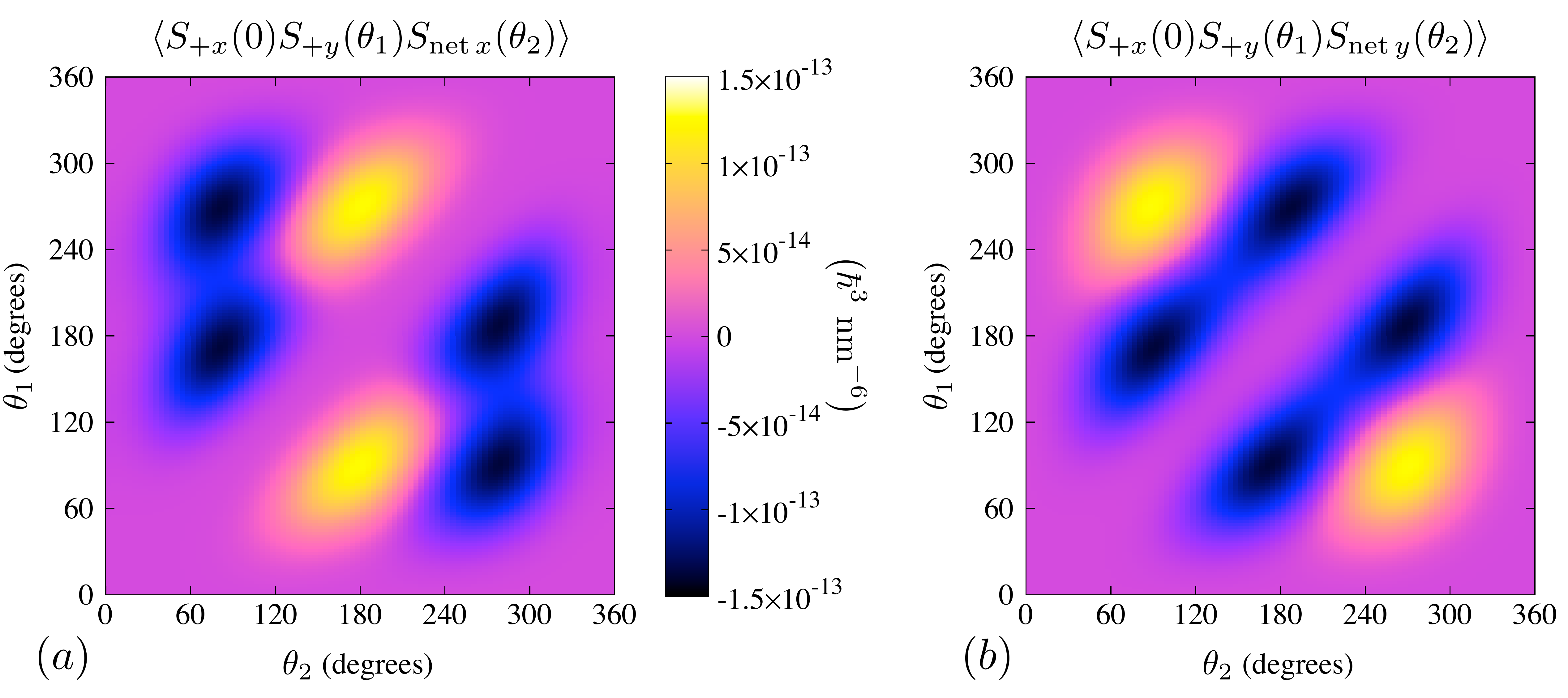}}
  \caption{\label{fig:3PtGS_tt} (Color online.) Three-point spin
    correlations along the ring of maximum single-particle density
    $r_{\text{max}}$ = 48 nm for the four-particle ground state.
    Specifically, the net $S_x$ (a) and net $S_y$ (b) are determined
    with respect to a particle of spin-up projection along the
    $x$-axis at $\vect{r}_0 = (r_{\text{max}},\, 0)$ and a particle of
    spin-up projection along the $y$-axis at $\vect{r}_1 =
    (r_{\text{max}},\theta_1)$.}
\end{figure}

\begin{figure}
  \centering
  \resizebox{\fig}{!}{\includegraphics{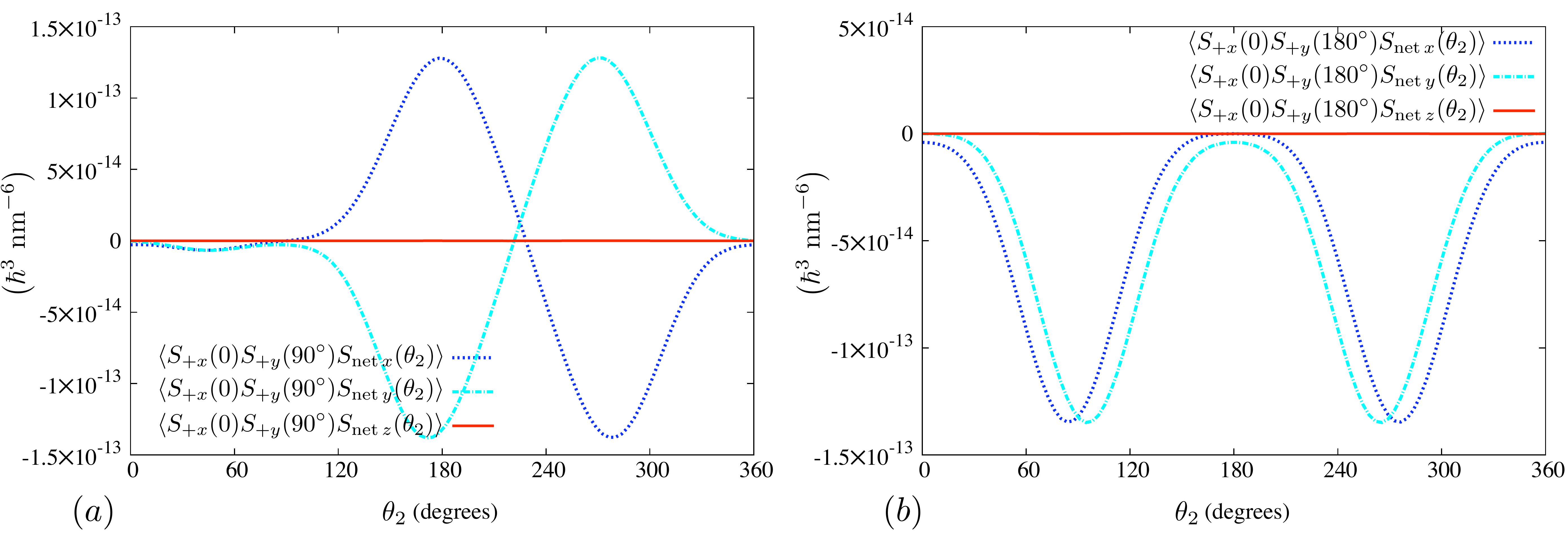}}
  \caption{\label{fig:3PtGS} (Color online.)  Three-point spin
    correlations along the ring of maximum single-particle density
    $r_{\text{max}} = 48$~nm in the four-particle ground-state
    manifold with respect to a particle that is spin-up along the
    $x$-axis at $\vect{r}_0 = (r_{\text{max}},\, 0)$ and a particle
    that is spin-up along the $y$-axis at $\vect{r}_1 =
    (r_{\text{max}}, 90^\circ)$ (a) and $\vect{r}_1 = (r_{\text{max}},
    180^\circ)$ (b).}
\end{figure}

To investigate possible chiral textures, we examine three-point spin
correlations in the four-particle ground-state manifold with respect
to one particle spin-up along the $x$-axis at $\vect{r}_0$, and a
second particle spin-up along the $y$-axis at $\vect{r}_1 \equiv
(r_{\text{max}}, \theta_1)$.  Figure~\ref{fig:3PtGS_tt}(a) shows the
net $x$ spin distribution and Fig.~\ref{fig:3PtGS_tt}(b) shows the net
$y$ spin distribution in the system at every angular position
$\theta_2$ as a function of $\theta_1$.  (The net $z$ correlations are
negligible and not shown.)
As $\theta_1$ approaches 90$^\circ$, two peaks (the two remaining
particles) emerge in the net $x$ and $y$ spin distributions at
$\theta_2 = 180^\circ$ and $\theta_2 = 270^\circ$.  As $\theta_1$
approaches 180$^\circ$, two different peaks arise at $\theta_2 =
90^\circ$ and $\theta_2 = 270^\circ$.  There is a third region, at
$\theta_1 = 270^\circ$ of large correlation that mirrors that at
$\theta_1 = 90^\circ$. Note the inversion symmetry through the point
$(\theta_1, \theta_2) = (180^\circ, 180^\circ)$.
Figure~\ref{fig:3PtGS} focuses on the two regions of large
correlation, when $\theta_1 = 90^\circ$ and 180$^\circ$, showing the
net $x$, $y$, and $z$ spin distributions as a function of $\theta_2$.
From these plots we conclude that chiral spin structures exist in the
ground-state manifold of the interacting four-particle system, but the
structures cannot be readily characterized by a definite winding about
any axis.

We now go on to consider the lowest-energy excitation above this
ground-state manifold, where we $do$ uncover winding textures.

\subsection{First Excited State}
\label{sec:N43PtES1}
The first excited state in the four-particle system is non-degenerate
with quantum numbers $L_z = 0$, $S = 0$.  It too has a circularly
symmetric single-particle density distribution about the origin, with
a peak at $r_{\text{max}} \equiv 48$~nm.

\begin{figure}
  \centering
  \resizebox{\fig}{!}{\includegraphics{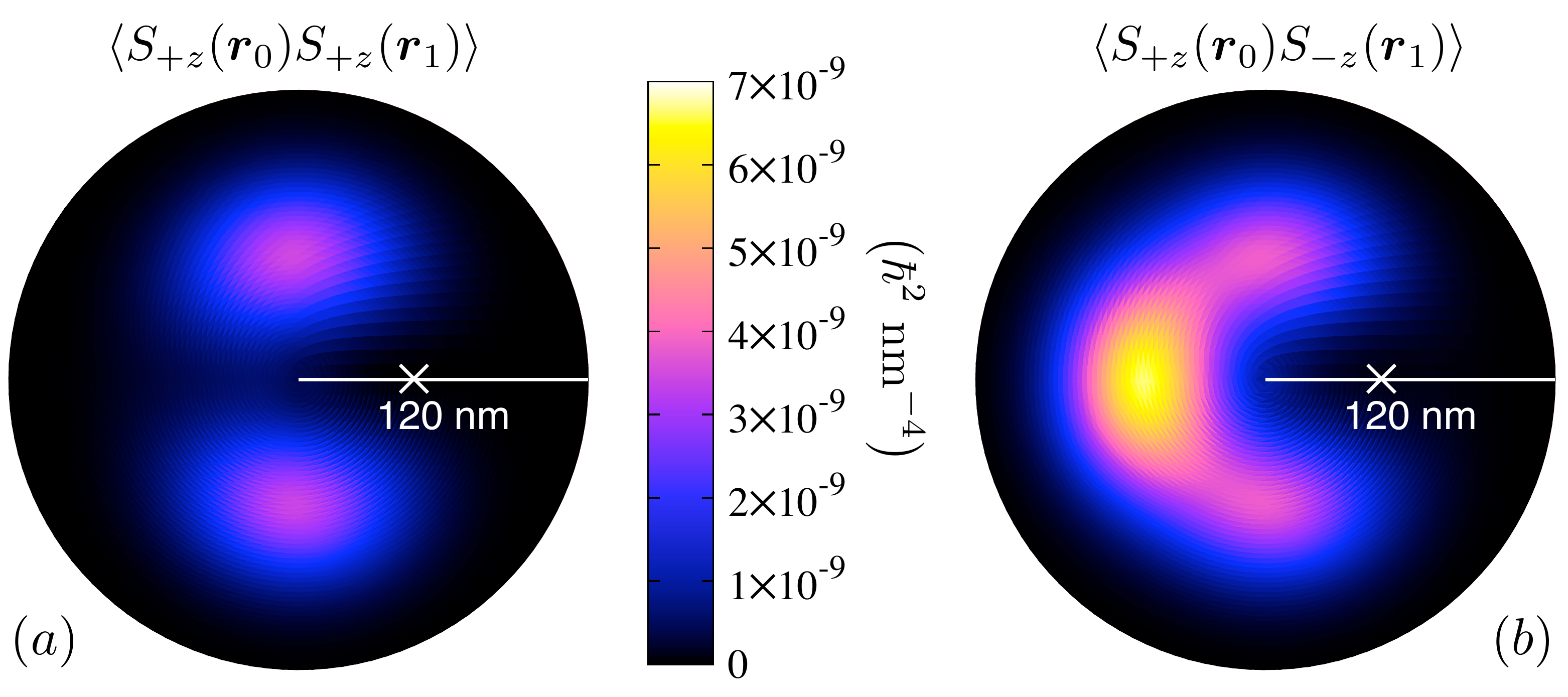}}
  \caption{\label{fig:N42pt1ES} (Color online.)  Two-point spin
    correlations in the first excited state of the four-particle
    system with respect to a spin-up particle at $\vect{r}_0 =
    (48~\text{nm},\, 0)$ (white cross).  Shown are the spin-up (a) and
    spin-down (b) distributions.}
\end{figure}

Figure~\ref{fig:N42pt1ES} shows the two-point spin correlations
throughout the plane of the dot with respect to a spin-up particle at
$\vect{r}_0 \equiv (r_{\text{max}},\, 0)$.  The distribution of the
spin-up and spin-down densities are shown.
The spin-up density shown in Fig.~\ref{fig:N42pt1ES}(a) contains two
peaks at $\theta_1 = 90^\circ$ and $\theta_1 = 270^\circ$, both along
the ring $r_{\text{max}}$.  The probability drops to approximately one
tenth of its magnitude between the peaks, at $\theta_1 = 180^\circ$.
The spin density goes to zero as $\vect{r}_1$ approaches $\vect{r}_0$,
giving evidence for a Pauli vortex at $\vect{r}_0$.  The spin-down
density shown in Fig.~\ref{fig:N42pt1ES}(b) has three peaks, the
largest at $\theta_1 = 180^\circ$, and two smaller ones of equal
magnitude at $\theta_1 = 90^\circ$ and $\theta_1 = 270^\circ$. In
contrast to the ground state (see Fig.~\ref{fig:N42PtGS}), this
manifold does not exhibit antiferromagnetic order.

\begin{figure}
  \centering
  \resizebox{\fig}{!}{\includegraphics{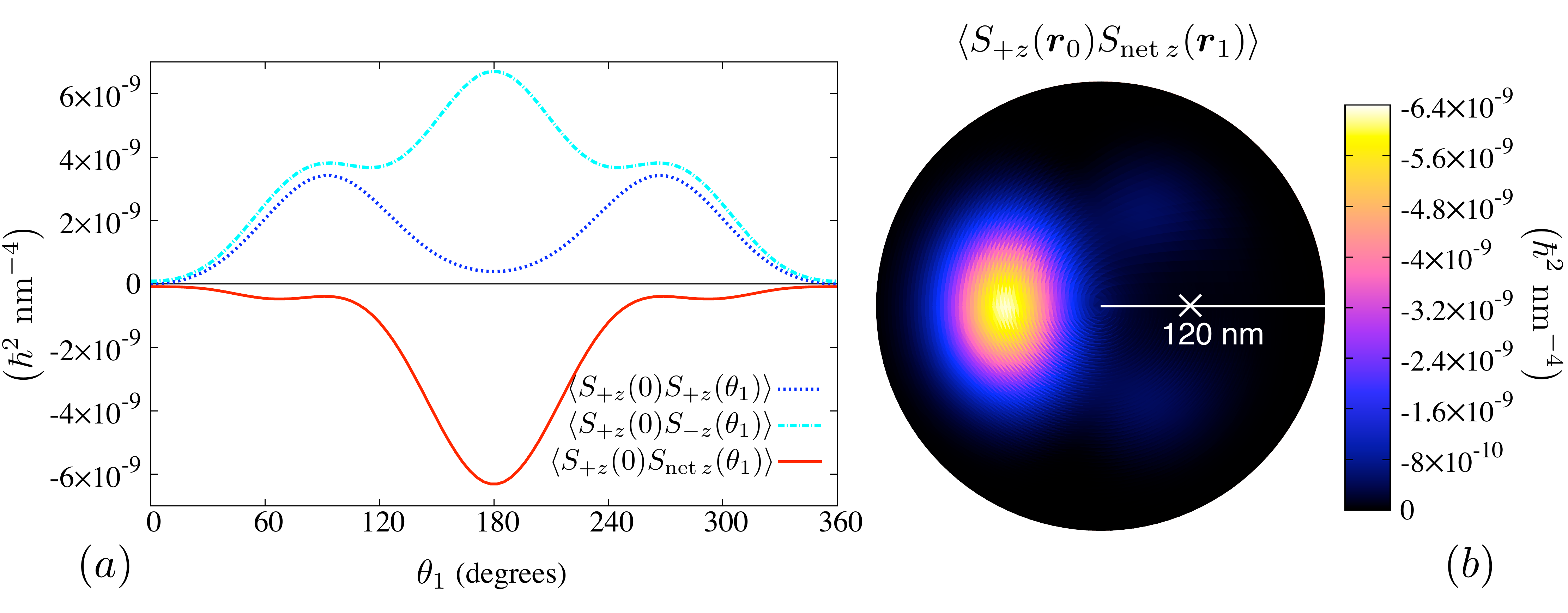}}
  \caption{\label{fig:N42pt1ES_net} (Color online.)  (a) Trace of the
    two-point spin distribution along the ring $r_{\text{max}} =
    48$~nm with respect to a spin-up particle at $\vect{r}_0 =
    (48~\text{nm},\, 0)$ in the four-particle first excited state.
    (b) Net two-point distribution of $S_z$ throughout the QD for the
    same state.}
\end{figure}

\begin{figure}
  \centering
  \resizebox{\smallfig}{!}{\includegraphics{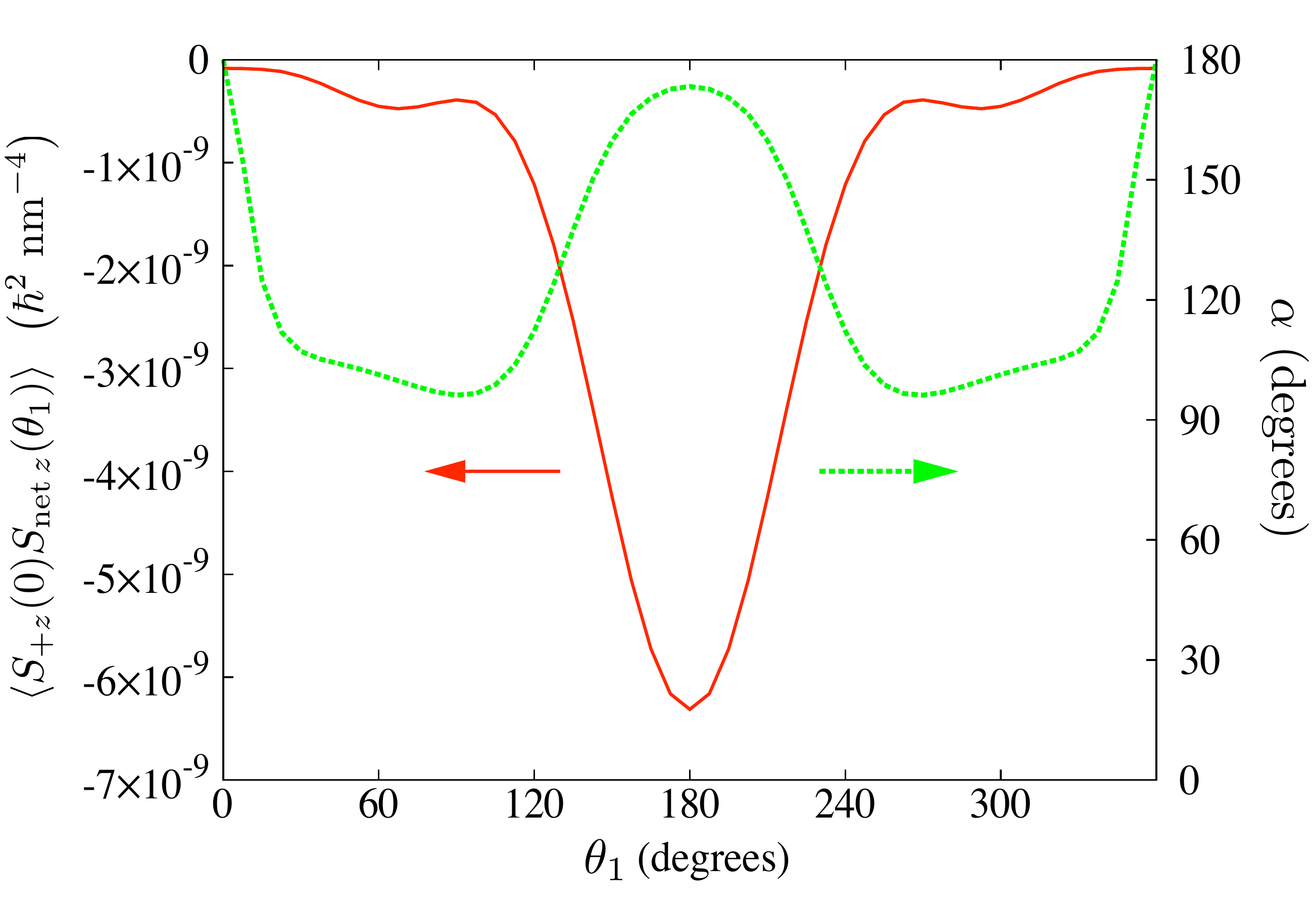}}
  \caption{\label{fig:N4S1canting} (Color online.) Canting angle
    $\alpha$ as a function of $\theta_1$ along the ring
    $r_{\text{max}} = 48$~nm in the four-particle first excited state.
    The net spin density from the two-point calculation along the same
    ring $r_{\text{max}}$ is shown for reference.}
\end{figure}

The two smaller peaks in the spin-down distribution are approximately
half the magnitude of the large peak and approximately equal to the
magnitude of the two peaks in the spin-up distribution.  The net $S_z$
distribution at $\theta_1 = 90^\circ$ and $\theta_1 = 270^\circ$ is
therefore approximately zero, indicating an in-plane orientation.
This is further evident in the trace along the ring $r_{\text{max}}$
shown in Fig.~\ref{fig:N42pt1ES_net}(a), and the net spin distribution
shown in Fig.~\ref{fig:N42pt1ES_net}(b).
The single peak at $\theta_1 = 180^\circ$ is composed primarily of a
single spin-species.  These results are consistent with two-point spin
correlations in Ref.~\onlinecite{Ghosal2007}.  Of note here is that,
in the transition from spin-up at $\theta_1 = 0^\circ$ to spin-down at
$\theta_1 = 180^\circ$, the spin lies almost completely in the plane
of the QD.  Calculation of the canting angle along the ring
$r_{\text{max}}$ as a function of $\theta_1$ is plotted in
Fig.~\ref{fig:N4S1canting}.
In this case there are two locations where the net spin has the
greatest canting.  Relative to the spin-up particle at $\vect{r}_0 =
(r_{\text{max}},\, 0)$, as $\theta_1$ approaches $\theta_0$ = 0 and
also as $\theta_1$ approaches 180$^\circ$, the canting angle
approaches 180$^\circ$. In the regions along $r_{\text{max}}$ where
the net spin distribution is almost zero, the canting angle approaches
90$^\circ$, showing that the net spin exists in the x-y plane.
However, the two-point correlation function is incapable of
determining the orientation of the spin within the plane.  We
therefore turn to the three-point correlation function in order to
determine the orientation of the in-plane spin component as it
transitions between spin-up and spin-down.

\begin{figure}
  \centering
  \resizebox{\fig}{!}{\includegraphics{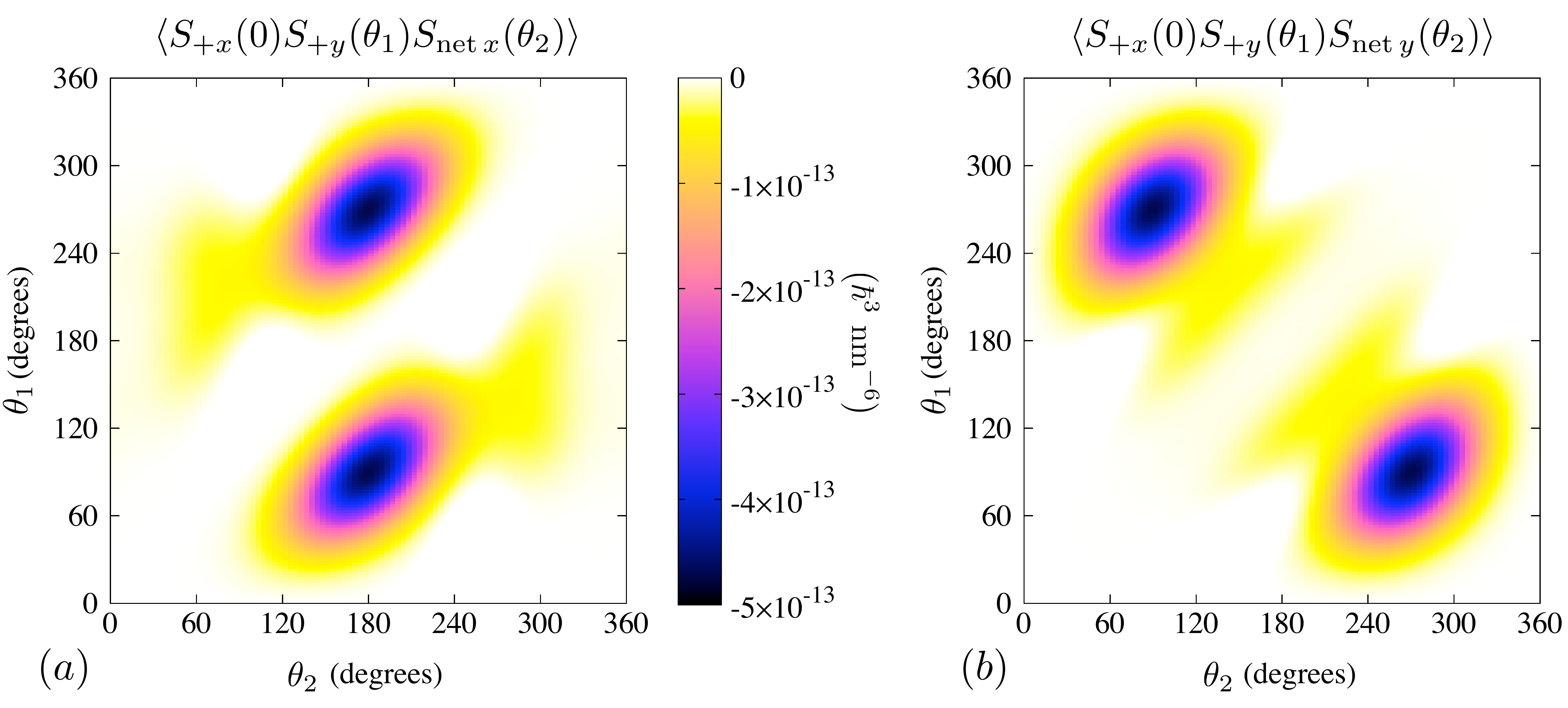}}
  \caption{\label{fig:3PtES1_tt} (Color online.)  Three-point
    correlations along the ring $r_{\text{max}} = 48$~nm in the
    four-particle first excited state with respect to a spin-up
    particle along the $x$-axis at $\vect{r}_0 = (r_{\text{max}},\,
    0)$ and a spin-up particle along the $y$-axis at $\vect{r}_1 =
    (r_{\text{max}},\,\theta_1)$.  Shown are the spin distributions
    along $\vect{r}_2 = (r_{\text{max}},\,\theta_2)$ for the net spin
    along the $x$-axis (a) and along the $y$-axis (b).}
\end{figure}

\begin{figure}
  \centering
  \resizebox{\fig}{!}{\includegraphics{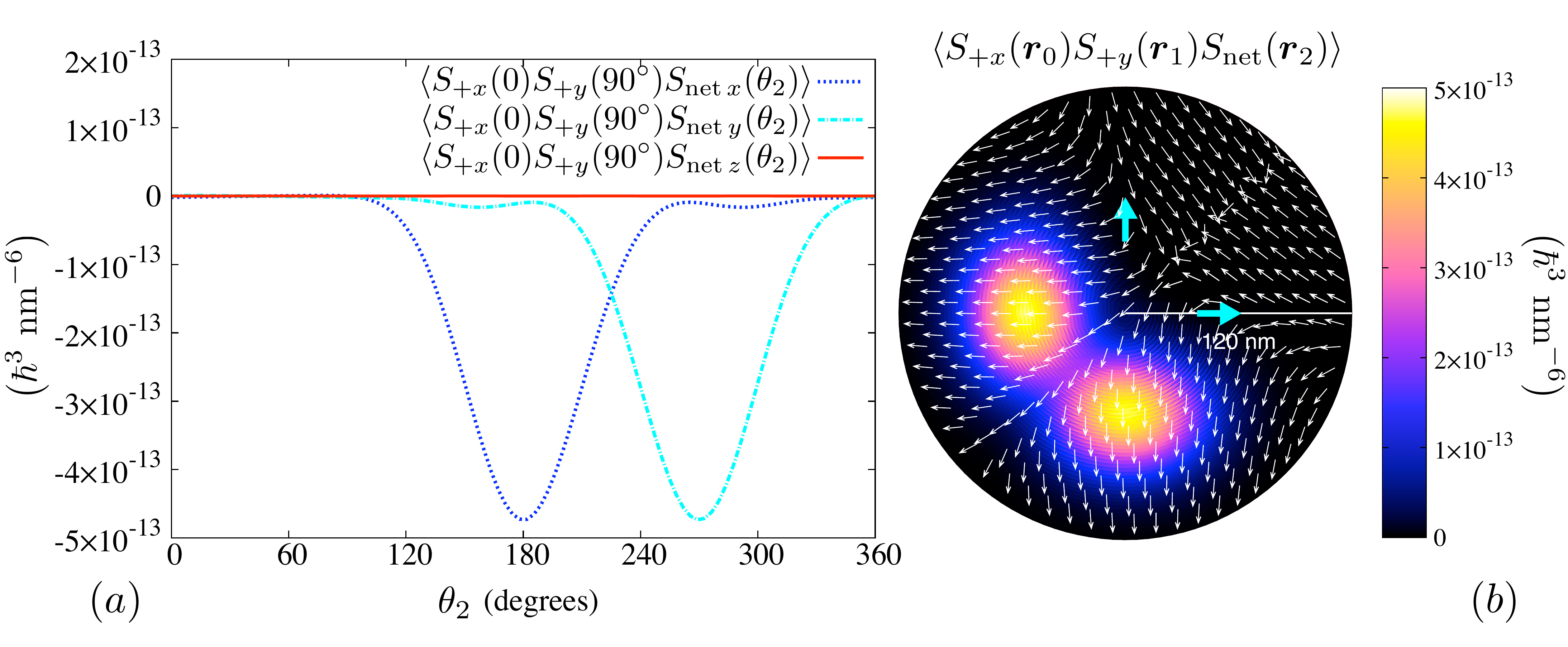}}
  \caption{\label{fig:3PtES1} (Color online.) (a) Trace along the ring
    $r_{\text{max}}$, revealing the net spin distributions in the
    four-particle first excited state given a spin-up particle along
    $x$ at $\vect{r}_0 =(r_{\text{max}},\, 0)$ and a spin-up particle
    along $y$ at $\vect{r}_1 =(r_{\text{max}},\, 90^\circ)$.  (b) The
    corresponding net spin distribution in the plane of the QD.  The
    vector field depicts the orientation and the color depicts the
    magnitude. The locations $\vect{r}_0$ and $\vect{r}_1$ are
    indicated by the blue arrows.}
\end{figure}

We calculate three-point spin correlations along the ring
$r_{\text{max}}$ with respect to a spin-up particle along the $x$-axis
at $\vect{r}_0 = (r_{\text{max}},\, 0)$, and a spin-up particle along
the $y$-axis at $\vect{r}_1 = (r_{\text{max}}, \theta_1)$.  The net
$x$ and net $y$ spin distributions are shown in
Fig.~\ref{fig:3PtES1_tt} as a function of $\theta_1$ and $\theta_2$.
These distributions reveal strong correlations at $\theta_1 =
90^\circ$ and $\theta_1 = 270^\circ$, consistent with the previous
two-point correlation.  Note the inversion symmetry about the point
$(\theta_1, \theta_2) = (180^\circ, 180^\circ)$.  The plots further
show that for $\theta_1 =90^\circ$, ($270^\circ$), the net spin is
predominantly spin-down along $x$ at $\theta_2 = 180^\circ$, and
predominantly spin-down along $y$ at $\theta_2 = 270^\circ$
($90^\circ$).  Focusing attention to the case of $\theta_1 =
90^\circ$, we plot in Fig.~\ref{fig:3PtES1}(a) the spin distribution
along the ring $r_{\text{max}}$ given a spin-up particle along $x$ at
$\theta_0 = 0^\circ$ and a spin-up particle along $y$ at $\theta_1 =
90^\circ$.  Figure~\ref{fig:3PtES1}(b) shows the effect of this
correlation on the net spin distribution throughout the plane of the
quantum dot.

These results for the net $x$ and $y$ spin distributions are
consistent with the $z$ distributions calculated by the two-point
correlations (see Fig.~\ref{fig:N42pt1ES_net}).  Those two-point
correlations revealed a single peak at $180^\circ$ indicating a spin
anti-aligned to the one at $\theta_0 = 0$ along the ring
$r_{\text{max}}$.  The three-point function with respect to a spin-up
particle along $x$ at $\vect{r}_0 =(r_{\text{max}},\, 0)$ and a
spin-up particle along $y$ at $\vect{r}_1 =(r_{\text{max}},\,
90^\circ)$ yields a net spin polarized along the negative $x$-axis at
$\theta = 180^\circ$, and a net spin polarized along the negative
$y$-axis at $\theta = 270^\circ$, on the ring $r_{\text{max}}$.  This
is evidence of a winding along the ring $r_{\text{max}}$.  From the
symmetry of our correlation operators, Eq.~\eqref{eq:3PtX_symm}, we
can deduce that there are in fact four orthogonal windings that wind
about the origin in this manner; two which begin with a particle
spin-up along the $x$-axis at $\vect{r}_0 =(r_{\text{max}},\, 0)$ and
then differ in the direction of their spin polarization along the
$y$-axis at $\vect{r}_1 =(r_{\text{max}},\,90^\circ)$ (i.e. in the
chirality of the rotation), and two which begin with a particle
spin-down along the $x$-axis at $\vect{r}_0 =(r_{\text{max}},\, 0)$,
and again differ in their chirality along the ring $r_{\text{max}}$.
Thus, we can characterize the first excited state of the four-particle
droplet as a superposition of four different windings, differing by
their chirality (clockwise and counter clockwise), and by a topological
charge ($\pm 1$).  (See Ref.~\onlinecite{loss:1996:berry_phase} for a
thorough semiclassical description).

\section{Summary}
\label{sec:summary}

We have computed spin correlations in the two lowest-lying states of
three-particle and four-particle circular two-dimensional QDs to
resolve spin textures that exist in the system in the presence of
strong, long-range Coulomb repulsion at zero magnetic field.  Our
findings are summarized in Table~\ref{tab:summary}.
\begin{table*}
  \caption{\label{tab:summary} Summary of textures uncovered in the
    respective degenerate manifolds.}
  \centering
  \begin{ruledtabular}
    \begin{tabular}{lclll}
      Manifold & Single Particle Densities & 2 Point & 3 Point & Canting \\ \hline
      3-particle ground state & $r_{max} = 39$ nm & triangular lattice & splayed & 131$^\circ$$^*$\\
      3-particle first excited state & $r_{max} = 41$ nm & triangular lattice & splayed & 53$^\circ$\\
      4-particle ground state & $r_{max} = 48$ nm & antiferromagnetic order & no winding & varies \\
      4-particle excited state & $r_{max} = 48$ nm & in-plane configuration & in-plane winding & varies
    \end{tabular}
  \end{ruledtabular}
  \raggedleft $^*$(\emph{at peaks})
\end{table*}
From the one-point correlation, we determine the annular regions of
maximum spin-density in the QD.  As expected, the radial distance of
this region from the origin depends on the number of confined
particles, and on the strength of the Coulomb repulsion.

We further compute two-point spin correlation functions to determine
the correlations in the spin field along a given direction.  Each
resulting spin distribution is symmetric through the diameter of the
QD on which one particle is located.  The details of the spin
configurations are dependent on the unique quantum characteristics of
the state, particularly the spin and orbital angular momentum quantum
numbers.  The two-point correlation calculations for each state
suggest the presence of a Pauli vortex at the position of the fixed
particle.  In addition, these results reveal an incipient Wigner phase
in the states examined above.

To uncover chiral textures, the three-point spin correlations are
calculated.  We plot the spin density with respect to two particles
with mutually perpendicular spins.  The three-particle states we have
investigated exhibit splayed textures.  The four-particle ground state
exhibits an antiferromagnetic texture, and the first excited state
exhibits winding textures.  Here, the results indicate that, given a
particle with spin at a point in the dot, the spin field rotates
through a plane perpendicular to the original spin orientation with a
full $2 \pi$ winding as one moves along a closed trajectory about the
origin.  Importantly, given the finite size of the system and the full
O(3) spin symmetry present, these winding textures are only
quasi-topological in character.  The rotational spin symmetry, in
particular, allows the spins to continually deform into the trivial
texture (or, rather, to the ground-state texture).  In a larger,
semiclassical system with at least uniaxial anisotropy, the analogous
textures \emph{will} have topological character.  In fact, in 2DEGs
and bulk systems, skyrmions and other spin textures have been shown to
exist.\cite{Kumada2006, Khandelwal2001, Gervais2005, Hsieh2009,
  Muehlbauer2009}

In the present small but fully quantum system, incipient topological
structures may manifest themselves if a coupling were introduced
between the spin field and the spatial orientation of the quantum dot
itself, analogous to the anisotropies typically found in much larger
magnetic systems.  This could come about through spin-orbit coupling.
There is evidence that such a coupling may break the degeneracy
between the spin-winding states.\cite{Stevenson:Unpublished} Together
with the strong Coulomb repulsion present in these small quantum
systems, the chiral structures that emerge should exhibit longer
lifetimes and lower decoherence rates than their more conventional
counterparts.

Improvements in the lifetime of these states could be determined, for
example, by comparing the relaxation rates of the states at and away
from their degeneracy point by use of single-shot
measurements.\cite{Hanson2005, Barthel2009} Experimental methods exist
for differentiating between fully polarized spin states and correlated
spin texture states in the ground state of a QD by measuring the
excitation spectra of the QD as a function of magnetic
field.\cite{Ciorga2003, Sachrajda2004, Nishi2006} These methods can be
used to determine when a spin texture state is present in the system.
The issues of spin-orbit coupling and state lifetimes will be
investigated in future work.

\begin{acknowledgments}
  This work was supported by the Natural Science and Engineering
  Research Council of Canada, and by the Lockheed Martin Corporation.
\end{acknowledgments}

\appendix
\setcounter{section}{-1}
\section{Derivation of Product Spin Operators}
\label{app:spinops}
\setcounter{section}{+1} In this appendix we discuss the general
properties of the product-operators used in our above analysis, as
well as the details associated with the derivation of each
product-operator.

\subsection{One-Body Spin Operators}
\label{sec:AppOneBody}
A general one-body operator is expressed in canonical second-quantized
form as
\begin{equation}
  \label{eq:oneBodyOp}
  \hat{U} = \sum_{\alpha\,\beta}U_{\alpha \beta}\, \hat{c}_\alpha^\dagger \, \hat{c}^{\phantom{\dagger}}_\beta,
\end{equation}
where $U_{\alpha \beta}=\rbra{\alpha}\hat{U}\rket{\beta}$ is a matrix
element of the operator,\cite{Negele1988} and
$\hat{c}_{\alpha}^\dagger$ and $\hat{c}^{\phantom{\dagger}}_{\alpha}$
are second-quantized Fermi operators, respectively creating and
destroying a particle in state $|\alpha\rangle$.

For a system of spin-1/2 fermions, the operator for the spin density
at position $\vect{r}$ is given by ($\hbar = 1$)
\begin{equation}
  \label{eq:spinDensity}
  \widehat{\vect{S}}(\vect{r}) =
  \frac{1}{2}\sum_{ss'} \hat{\psi}_s^\dagger(\vect{r})
  \vect{\widehat{\sigma}}^{ss'} \hat{\psi}_{s'}(\vect{r}),
\end{equation}
where $\vect{\widehat{\sigma}} = (\hat{\sigma}_x, \hat{\sigma}_y,
\hat{\sigma}_z)$ are the Pauli spin matrices and
$\hat{\psi}_s^\dagger(\vect{r})$ and $\hat{\psi}_{s}(\vect{r})$ are
the field operators.

Equation~(\ref{eq:spinDensity}) yields the net spin density at point
$\vect{r}$.  We are additionally interested in distinguishing the
spin-up and spin-down densities along each coordinate-axis.  We define
a general set of spin operators $\hat{S}_{\pm\alpha}(\vect{r})$ that
separately determines the spin-up and spin-down densities along
$\alpha = x, y, z$ at position $\vect{r}$.  In the $S_z$ basis, the
operator for the $\pm x$ spin density, for example, is given by
\begin{multline} 
  \label{eq:SPlusMinusX}
  \hat{S}_{\pm x}(\vect{r}) = \frac{1}{4} \left[ \left(
      \hat{\psi}_\uparrow^\dagger(\vect{r})
      \hat{\psi}_\uparrow^{\phantom{\dagger}}(\vect{r}) + 
      \hat{\psi}_\downarrow^\dagger(\vect{r})
      \hat{\psi}_\downarrow^{\phantom{\dagger}}(\vect{r}) \right) \right.\\
  \pm \left. \left(
      \hat{\psi}_\uparrow^\dagger(\vect{r})
      \hat{\psi}_\downarrow^{\phantom{\dagger}}(\vect{r}) + 
      \hat{\psi}_\downarrow^\dagger(\vect{r})
      \hat{\psi}_\uparrow^{\phantom{\dagger}}(\vect{r}) \right) \right].
\end{multline}
This operator can be derived from the field operators,
$\hat{\psi}_{\pm x}^\dagger(\vect{r}) =
[\hat{\psi}_\uparrow^\dagger(\vect{r}) \pm
\hat{\psi}_\downarrow^\dagger(\vect{r})] / \sqrt{2}$, and $\hat{S}_{\pm
  x}(\vect{r}) = \hat{\psi}_{\pm x}^\dagger(\vect{r}) \hat{\psi}_{\pm
  x}(\vect{r}) / 2$.
Similarly, the operators along the other two orthogonal directions are
\begin{multline} 
  \label{eq:SPlusMinusY}
  \hat{S}_{\pm y}(\vect{r}) = \frac{1}{4}\left[ \left(
      \hat{\psi}_\uparrow^\dagger(\vect{r})
      \hat{\psi}_\uparrow^{\phantom{\dagger}}(\vect{r}) +
      \hat{\psi}_\downarrow^\dagger(\vect{r})
      \hat{\psi}_\downarrow^{\phantom{\dagger}}(\vect{r}) \right) \right. \\
  \pm \left. i \left(
      \hat{\psi}_\downarrow^\dagger(\vect{r})
      \hat{\psi}_\uparrow^{\phantom{\dagger}}(\vect{r}) -
      \hat{\psi}_\uparrow^\dagger(\vect{r})
      \hat{\psi}_\downarrow^{\phantom{\dagger}}(\vect{r}) \right) \right],
\end{multline}
and
\begin{equation}
  \label{eq:SPlusZMinusZ}
  \hat{S}_{+z}(\vect{r}) = \frac{1}{2} 
  \hat{\psi}_\uparrow^\dagger(\vect{r})
  \hat{\psi}_\uparrow^{\phantom{\dagger}}(\vect{r}),
  \quad
  \hat{S}_{-z}(\vect{r}) = \frac{1}{2} 
  \hat{\psi}_\downarrow^\dagger(\vect{r})
  \hat{\psi}_\downarrow^{\phantom{\dagger}}(\vect{r}).
\end{equation}
By defining the net spin along an axis to be the difference between
the spin-up and the spin-down density along that same axis
($\hat{S}_x(\vect{r}) = \hat{S}_{+x}(\vect{r}) -
\hat{S}_{-x}(\vect{r})$, etc.), we obtain the components of
Eq.~\eqref{eq:spinDensity}.  Upon integrating these net spin
components over all space, we recover the usual expressions $\hat{S}_x
= (\hat{S}_+ + \hat{S}_-) / 2$, $\hat{S}_y = (\hat{S}_+ - \hat{S}_-) /
(2i)$, and $\hat{S}_z = (\hat{S}_\uparrow - \hat{S}_\downarrow) / 2$,
where $\hat{S}_+$ and $\hat{S}_-$ are $S_z$ raising and lowering
operators, respectively.

The operators in Eqs.~(\ref{eq:SPlusMinusX}, \ref{eq:SPlusMinusY})
each contain terms which flip spins.  But since the Hamiltonian,
Eq.~(\ref{eq:hamil}), conserves spin, and since we include spin
quantum numbers to classify our states, these terms give zero
contribution to Eq.~(\ref{eq:expectRho}) for one-body spin operators.
As a consequence of spin conservation we have for any degenerate
manifold at zero magnetic field, $\langle S_{x}(\vect{r}) \rangle =
\langle S_{y}(\vect{r}) \rangle = \langle S_{z}(\vect{r}) \rangle =
0$.  Along the quantization ($z$) axis, we can distinguish between the
spin-up and the spin-down density at position $\vect{r}$, but we
cannot distinguish between the spin-up and spin-down density along the
two orthogonal directions: Formally, $\langle S_{+x}(\vect{r}) \rangle
= \langle S_{-x}(\vect{r}) \rangle$, and similarly for the $y$
direction.  Consequently, it suffices to calculate only the one-point
correlations of $\hat{S}_{\pm z}(\vect{r})$, as in
Eq.~\eqref{eq:SPlusMinusZCanonical}.

\subsection{Two-Body Spin Operators}
\label{sec:AppTwoBody}
To investigate correlations along a particular axis, the two-point
correlations functions are required.  On physical grounds, we require
that operators be symmetric in their indexes.\cite{Negele1988} For
one-body operators, we require that $U_{\alpha \beta} = U_{\beta
  \alpha}$.  For a two-body operator, we similarly require that
$V_{\alpha \beta \gamma \delta} = V_{\beta \alpha \delta \gamma}$.  In
general, a product of $N$ one-body operators is not an $N$-body
operator.  Let $\hat{U}$ and $\hat{V}$ each be a one-body operator.
Their product can be written as
\begin{multline} 
  \label{eq:twoFactor}
  \hat{U} \hat{V} = \sum_{ij}(UV)_{ij}\,\hat{c}_{i}^\dagger
  \hat{c}_{j}^{\phantom{\dagger}}
  \\
  + \frac{1}{2} \sum_{ijkl}\left( U_{ik}V_{jl} + U_{jl}V_{ik} \right)
  \hat{c}_{i}^\dagger\hat{c}_{j}^\dagger 
  \hat{c}_{l}^{\phantom{\dagger}} \hat{c}_{k}^{\phantom{\dagger}},
\end{multline}
where $(UV)_{ij} \equiv \rbra{i} \hat{U} \hat{V} \rket{j}$,
$U_{ik}V_{jl} \equiv \rbra{i}\hat{U}\rket{k}\rbra{j}\hat{V}\rket{l}$.
The product of two one-body operators is in fact a sum of canonical
one-body and two-body operators.

There is no preferred axis along which to calculate our two-point
correlations because the net spin in each degenerate manifold is zero,
but one may exploit the symmetry; for example, the correlation between
a spin-up particle and the remaining spin-up distribution is identical
to the correlation between a spin-down particle and the remaining
spin-down distribution.  In general, the correlations between a
particle with spin and the remaining particles of parallel spin will
be the same for any orientation, as will the correlations between a
particle with spin and the remaining particles of antiparallel-spin.
Due to spin conservation, two-point correlations between particles
with perpendicular spin do not provide additional correlation
information and are therefore not considered.

The two-point correlations investigated above are $\langle
S_{+z}(\vect{r}_0) S_{+z}(\vect{r}_1) \rangle$ and $\langle
S_{+z}(\vect{r}_0) S_{-z}(\vect{r}_1) \rangle$ in
Eq.~\eqref{eq:twoPointOp}.  With the condition $\vect{r}_0 \neq
\vect{r}_1$ (while still allowing $\vect{r}_0 \rightarrow
\vect{r}_1$), the one-body term vanishes.  As described in
Sec.~\ref{sec:AppOneBody}, spin symmetry implies that all four of
$\langle S_{+z}(\vect{r}_0) S_{\pm x}(\vect{r}_1) \rangle$ and
$\langle S_{+z}(\vect{r}_0) S_{\pm y}(\vect{r}_1) \rangle$ are
identical and equal to $\langle ( S_{+z}(\vect{r}_0)
S_{+z}(\vect{r}_1) + S_{+z}(\vect{r}_0) S_{-z}(\vect{r}_1) ) \rangle /
2$.

\subsection{Three-Body Spin Operators}
\label{sec:AppThreeBody}
To probe for chiral textures, we compute the three-point correlation
functions.  These are a product of three one-body operators.  In
canonical form, the product of three one-body operators is the sum of
a canonical one-body, two-body, and three-body operator,
\begin{widetext}
  \begin{multline}
    \label{eq:threeFactor}
    \hat{U} \hat{V} \hat{W}  =
    \sum_{i\,j}(UVW)_{ij} \, \hat{c}_{i}^\dagger \hat{c}_{j}
    + \frac{1}{2} \sum_{ijkl} 
    \Bigl( (UV)_{ik}W_{jl} + U_{ik}(VW)_{kl} + (UW)_{ik}V_{kl}
    \\ 
    \shoveright{ \mbox{}
      + (UV)_{jl}W_{ik} +U_{jl}(VW)_{ik} + (UW)_{jl}V_{ik} \Bigr) 
      \hat{c}_{i}^\dagger\hat{c}_{j}^\dagger \hat{c}_{l}^{\phantom{\dagger}}
      \hat{c}_{k}^{\phantom{\dagger}}}
    \\  
    \mbox{} + \frac{1}{3!} \sum_{\substack{ijk\\lmn}}
    \Bigl(U_{il}V_{jm}W_{kn} + U_{il}V_{kn}W_{jm} +
    U_{jm}V_{il}W_{kn} 
    \\ \mbox{}
    + U_{jm}V_{kn}W_{il} + U_{kn}V_{il}W_{jm} +
    U_{kn}V_{jm}W_{il}\Bigr) 
    \hat{c}_{i}^\dagger \hat{c}_{j}^\dagger \hat{c}_{k}^\dagger
    \hat{c}^{\phantom{\dagger}}_{n} 
    \hat{c}^{\phantom{\dagger}}_{m} \hat{c}^{\phantom{\dagger}}_{l},
  \end{multline}
\end{widetext}
where $(UVW)_{ij} = \rbra{i} \hat{U} \hat{V} \hat{W} \rket{j}$.  Each
of the matrix elements in Eq.~(\ref{eq:threeFactor}) is symmetric
under appropriate interchange of indexes; for a three-body matrix
element $V_{ijklmn}$, for example, we require $V_{ijklmn} =
V_{ikjlnm}$, and so on.

We express our three-point correlation operators in the symmetric form
shown in Eq.~\eqref{eq:threeFactor}.  As in Sec~\ref{sec:AppTwoBody},
spin symmetry implies that the one-body and two-body pieces of
Eq.~(\ref{eq:threeFactor}) vanish when $\vect{r}_0$, $\vect{r}_1$, and
$\vect{r}_2$ are all different.  Because our averages are taken with
respect to spin-conserving states, we need only consider terms in the
correlation function which themselves conserve spin.  We can thus
write the product operators as they are given in
Eq.~\eqref{eq:3PtCanonical}.

\bibliography{Stevenson2010_v2}

\begin{thebibliography}{53}
\expandafter\ifx\csname natexlab\endcsname\relax\def\natexlab#1{#1}\fi
\expandafter\ifx\csname bibnamefont\endcsname\relax
  \def\bibnamefont#1{#1}\fi
\expandafter\ifx\csname bibfnamefont\endcsname\relax
  \def\bibfnamefont#1{#1}\fi
\expandafter\ifx\csname citenamefont\endcsname\relax
  \def\citenamefont#1{#1}\fi
\expandafter\ifx\csname url\endcsname\relax
  \def\url#1{\texttt{#1}}\fi
\expandafter\ifx\csname urlprefix\endcsname\relax\def\urlprefix{URL }\fi
\providecommand{\bibinfo}[2]{#2}
\providecommand{\eprint}[2][]{\url{#2}}

\bibitem[{\citenamefont{Kouwenhoven et~al.}(2001)\citenamefont{Kouwenhoven,
  Austing, and Tarucha}}]{kouwen01:few.elect.quant.dots}
\bibinfo{author}{\bibfnamefont{L.~P.} \bibnamefont{Kouwenhoven}},
  \bibinfo{author}{\bibfnamefont{D.~G.} \bibnamefont{Austing}},
  \bibnamefont{and} \bibinfo{author}{\bibfnamefont{S.}~\bibnamefont{Tarucha}},
  \bibinfo{journal}{Rep.\ Prog.\ Phys.} \textbf{\bibinfo{volume}{64}},
  \bibinfo{pages}{701} (\bibinfo{year}{2001}).

\bibitem[{\citenamefont{Bukowski and Simmons}(2002)}]{bukow02:quant.dot.resear}
\bibinfo{author}{\bibfnamefont{T.~J.} \bibnamefont{Bukowski}} \bibnamefont{and}
  \bibinfo{author}{\bibfnamefont{J.~H.} \bibnamefont{Simmons}},
  \bibinfo{journal}{Crit.\ Rev.\ Solid State} \textbf{\bibinfo{volume}{27}},
  \bibinfo{pages}{119} (\bibinfo{year}{2002}).

\bibitem[{\citenamefont{Reimann and
  Manninen}(2002)}]{reiman02:elect.struc.quant.dots}
\bibinfo{author}{\bibfnamefont{S.~M.} \bibnamefont{Reimann}} \bibnamefont{and}
  \bibinfo{author}{\bibfnamefont{M.}~\bibnamefont{Manninen}},
  \bibinfo{journal}{Rev.\ Mod.\ Phys.} \textbf{\bibinfo{volume}{74}},
  \bibinfo{pages}{1283} (\bibinfo{year}{2002}).

\bibitem[{\citenamefont{Burkard}(2006)}]{burkar06:theor.solid.state}
  \bibinfo{author}{\bibfnamefont{G.}~\bibnamefont{Burkard}},
  \bibinfo{note}{in}
  \emph{\bibinfo{title}{Handbook of Theoretical and Computational 
      Nanotechnology}},
  \bibinfo{note}{edited by}
  \bibinfo{editor}{\bibfnamefont{M.}~\bibnamefont{Rieth}} \bibnamefont{and}
  \bibinfo{editor}{\bibfnamefont{W.}~\bibnamefont{Schommers}}
  (\bibinfo{publisher}{American Scientific Publishers}, 
  \bibinfo{address}{Stevenson Ranch, CA}, \bibinfo{year}{2006}),
  \bibinfo{note}{Vol.} \bibinfo{volume}{3}.

\bibitem[{\citenamefont{Engel et~al.}(2004)\citenamefont{Engel, Kouwenhoven,
  Loss, and Marcus}}]{engel04:contr.spin.qubit}
\bibinfo{author}{\bibfnamefont{H.-A.} \bibnamefont{Engel}},
  \bibinfo{author}{\bibfnamefont{L.~P.} \bibnamefont{Kouwenhoven}},
  \bibinfo{author}{\bibfnamefont{D.}~\bibnamefont{Loss}}, \bibnamefont{and}
  \bibinfo{author}{\bibfnamefont{C.~M.} \bibnamefont{Marcus}},
  \bibinfo{journal}{Quantum Inf. Process.} \textbf{\bibinfo{volume}{3}},
  \bibinfo{pages}{115} (\bibinfo{year}{2004}), \eprint{cond-mat/0409294}.

\bibitem[{\citenamefont{Krenner et~al.}(2005)\citenamefont{Krenner, Stufler,
  Sabathil, Clark, Ester, Bichler, Abstreiter, Finley, and
  Zrenner}}]{Krenner05:Recent-advances}
\bibinfo{author}{\bibfnamefont{H.~J.} \bibnamefont{Krenner}},
  \bibinfo{author}{\bibfnamefont{S.}~\bibnamefont{Stufler}},
  \bibinfo{author}{\bibfnamefont{M.}~\bibnamefont{Sabathil}},
  \bibinfo{author}{\bibfnamefont{E.~C.} \bibnamefont{Clark}},
  \bibinfo{author}{\bibfnamefont{P.}~\bibnamefont{Ester}},
  \bibinfo{author}{\bibfnamefont{M.}~\bibnamefont{Bichler}},
  \bibinfo{author}{\bibfnamefont{G.}~\bibnamefont{Abstreiter}},
  \bibinfo{author}{\bibfnamefont{J.~J.} \bibnamefont{Finley}},
  \bibnamefont{and} \bibinfo{author}{\bibfnamefont{A.}~\bibnamefont{Zrenner}},
  \bibinfo{journal}{New J.~Phys.} \textbf{\bibinfo{volume}{7}},
  \bibinfo{pages}{184} (\bibinfo{year}{2005}), \eprint{cond-mat/0505731}.

\bibitem[{\citenamefont{Taylor et~al.}(2005)\citenamefont{Taylor, Engel, Dur,
  Yacoby, Marcus, Zoler, and Lukin}}]{Taylor2005}
\bibinfo{author}{\bibfnamefont{J.~M.} \bibnamefont{Taylor}},
  \bibinfo{author}{\bibfnamefont{H.-A.} \bibnamefont{Engel}},
  \bibinfo{author}{\bibfnamefont{W.}~\bibnamefont{D\"{u}r}},
  \bibinfo{author}{\bibfnamefont{A.}~\bibnamefont{Yacoby}}
  \bibinfo{author}{\bibfnamefont{C.~M.} \bibnamefont{Marcus}},
  \bibinfo{author}{\bibfnamefont{P.}~\bibnamefont{Zoller}}, \bibnamefont{and}
  \bibinfo{author}{\bibfnamefont{M.~D.} \bibnamefont{Lukin}},
  \bibinfo{journal}{Nature Physics} \textbf{\bibinfo{volume}{1}},
  \bibinfo{pages}{177} (\bibinfo{year}{2005}).

\bibitem[{\citenamefont{Foletti et~al.}(2009)\citenamefont{Foletti, Bluhm,
  Mahalu, Umansky, and Yacoby}}]{Foletti2009}
      \bibinfo{author}{\bibfnamefont{S.}~\bibnamefont{Foletti}},
      \bibinfo{author}{\bibfnamefont{H.}~\bibnamefont{Bluhm}},
      \bibinfo{author}{\bibfnamefont{D.}~\bibnamefont{Mahalu}},
      \bibinfo{author}{\bibfnamefont{V.}~\bibnamefont{Umansky}}, 
      \bibnamefont{and}
      \bibinfo{author}{\bibfnamefont{A.}~\bibnamefont{Yacoby}},
      \bibinfo{journal}{Nature Physics} \textbf{\bibinfo{volume}{5}},
      \bibinfo{pages}{903} (\bibinfo{year}{2009}).

\bibitem[{\citenamefont{Laird et~al.}(2010)\citenamefont{Laird, Taylor,
  DiVincenzo, Marcus, Hanson, and Gossard}}]{Laird2010}
      \bibinfo{author}{\bibfnamefont{E.~A.} \bibnamefont{Laird}},
      \bibinfo{author}{\bibfnamefont{J.~M.} \bibnamefont{Taylor}},
      \bibinfo{author}{\bibfnamefont{D.~P.} \bibnamefont{DiVincenzo}},
      \bibinfo{author}{\bibfnamefont{C.~M.} \bibnamefont{Marcus}},
      \bibinfo{author}{\bibfnamefont{M.~P.} \bibnamefont{Hanson}}, 
      \bibnamefont{and}
      \bibinfo{author}{\bibfnamefont{A.~C.} \bibnamefont{Gossard}},
      \bibinfo{journal}{Phys.\ Rev.\ B} \textbf{\bibinfo{volume}{82}},
      \bibinfo{pages}{075403} (\bibinfo{year}{2010}).

\bibitem[{\citenamefont{Ghosal et~al.}(2007)\citenamefont{Ghosal, Guclu,
  Umrigar, Ullmo, and Baranger}}]{Ghosal2007}
\bibinfo{author}{\bibfnamefont{A.}~\bibnamefont{Ghosal}},
  \bibinfo{author}{\bibfnamefont{A.~D.} \bibnamefont{Guclu}},
  \bibinfo{author}{\bibfnamefont{C.~J.} \bibnamefont{Umrigar}},
  \bibinfo{author}{\bibfnamefont{D.}~\bibnamefont{Ullmo}}, \bibnamefont{and}
  \bibinfo{author}{\bibfnamefont{H.~U.} \bibnamefont{Baranger}},
  \bibinfo{journal}{Phys.\ Rev.\ B} \textbf{\bibinfo{volume}{76}},
  \bibinfo{pages}{085341} (\bibinfo{year}{2007}).

\bibitem[{\citenamefont{Ghosal et~al.}(2006)\citenamefont{Ghosal, Guclu,
  Umrigar, Ullmo, and Baranger}}]{Ghosal2006}
\bibinfo{author}{\bibfnamefont{A.}~\bibnamefont{Ghosal}},
  \bibinfo{author}{\bibfnamefont{A.~D.} \bibnamefont{Guclu}},
  \bibinfo{author}{\bibfnamefont{C.~J.} \bibnamefont{Umrigar}},
  \bibinfo{author}{\bibfnamefont{D.}~\bibnamefont{Ullmo}}, \bibnamefont{and}
  \bibinfo{author}{\bibfnamefont{H.~U.} \bibnamefont{Baranger}},
  \bibinfo{journal}{Nature Physics} \textbf{\bibinfo{volume}{2}},
  \bibinfo{pages}{336} (\bibinfo{year}{2006}).

\bibitem[{\citenamefont{Loss and
  DiVincenzo}(1998)}]{Loss98:Quantum-computation}
\bibinfo{author}{\bibfnamefont{D.}~\bibnamefont{Loss}} \bibnamefont{and}
  \bibinfo{author}{\bibfnamefont{D.~P.} \bibnamefont{DiVincenzo}},
  \bibinfo{journal}{Phys.\ Rev.\ A} \textbf{\bibinfo{volume}{57}},
  \bibinfo{pages}{120} (\bibinfo{year}{1998}), \eprint{cond-mat/9701055}.

\bibitem[{\citenamefont{Burkard et~al.}(1999)\citenamefont{Burkard, Loss, and
  DiVincenzo}}]{Burkard99:Coupled-quantum}
\bibinfo{author}{\bibfnamefont{G.}~\bibnamefont{Burkard}},
  \bibinfo{author}{\bibfnamefont{D.}~\bibnamefont{Loss}}, \bibnamefont{and}
  \bibinfo{author}{\bibfnamefont{D.~P.} \bibnamefont{DiVincenzo}},
  \bibinfo{journal}{Phys.\ Rev.\ B} \textbf{\bibinfo{volume}{59}},
  \bibinfo{pages}{2070} (\bibinfo{year}{1999}), \eprint{cond-mat/9808026}.

\bibitem[{\citenamefont{Hanson et~al.}(2007)\citenamefont{Hanson, Kouwenhoven, 
      Petta, Tarucha, and Vandersypen}}]{Hanson2007}
\bibinfo{author}{\bibfnamefont{R.}~\bibnamefont{Hanson}},
  \bibinfo{author}{\bibfnamefont{L.~P.} \bibnamefont{Kouwenhoven}},
  \bibinfo{author}{\bibfnamefont{J.~R.} \bibnamefont{Petta}},
  \bibinfo{author}{\bibfnamefont{S}~\bibnamefont{Tarucha}}, \bibnamefont{and}
  \bibinfo{author}{\bibfnamefont{L.~M.~K.} \bibnamefont{Vandersypen}},
  \bibinfo{journal}{Rev. Mod. Phys.} \textbf{\bibinfo{volume}{79}},
  \bibinfo{pages}{1217} (\bibinfo{year}{2007}).

\bibitem[{\citenamefont{Ramsay}(2010)}]{Ramsay2010}
  \bibinfo{author}{\bibfnamefont{A.~J.}~\bibnamefont{Ramsay}},
  \bibinfo{journal}{Semicond. Sci. Technol.} \textbf{\bibinfo{volume}{25}},
  \bibinfo{pages}{103001} (\bibinfo{year}{2010}).

\bibitem[{\citenamefont{Averin and Goldman}(2001)}]{Averin2001}
\bibinfo{author}{\bibfnamefont{D.~V.} \bibnamefont{Averin}} \bibnamefont{and}
  \bibinfo{author}{\bibfnamefont{V.~J.} \bibnamefont{Goldman}},
  \bibinfo{journal}{Solid State Commun.} \textbf{\bibinfo{volume}{121}},
  \bibinfo{pages}{25} (\bibinfo{year}{2001}).

\bibitem[{\citenamefont{Kitaev}(2003)}]{Kitaev2003}
\bibinfo{author}{\bibfnamefont{A.~Y.} \bibnamefont{Kitaev}},
  \bibinfo{journal}{Annals of Physics} \textbf{\bibinfo{volume}{303}},
  \bibinfo{pages}{2} (\bibinfo{year}{2003}).

\bibitem[{\citenamefont{Das~Sarma et~al.}(2005)\citenamefont{Das~Sarma,
  Freedman, and Nayak}}]{Das-Sarma2005}
\bibinfo{author}{\bibfnamefont{S.}~\bibnamefont{Das~Sarma}},
  \bibinfo{author}{\bibfnamefont{M.}~\bibnamefont{Freedman}}, \bibnamefont{and}
  \bibinfo{author}{\bibfnamefont{C.}~\bibnamefont{Nayak}},
  \bibinfo{journal}{Phys.\ Rev.\ Lett.} \textbf{\bibinfo{volume}{94}},
  \bibinfo{pages}{166802} (\bibinfo{year}{2005}).

\bibitem[{\citenamefont{Bombin and Martin-Delgado}(2007)}]{Bombin2007}
\bibinfo{author}{\bibfnamefont{H.}~\bibnamefont{Bombin}} \bibnamefont{and}
  \bibinfo{author}{\bibfnamefont{M.~A.} \bibnamefont{Martin-Delgado}},
  \bibinfo{journal}{Phys.\ Rev.\ Lett.} \textbf{\bibinfo{volume}{98}},
  \bibinfo{pages}{160502} (\bibinfo{year}{2007}), ISSN \bibinfo{issn}{0031-9007
  (Print)}.

\bibitem[{\citenamefont{DiVincenzo}(2000)}]{DiVincenzo2000}
\bibinfo{author}{\bibfnamefont{D.~P.} \bibnamefont{DiVincenzo}},
  \bibinfo{journal}{Fortschr. Phys.} \textbf{\bibinfo{volume}{48}},
  \bibinfo{pages}{771} (\bibinfo{year}{2000}).

\bibitem[{\citenamefont{Duan et~al.}(2003)\citenamefont{Duan, Demler, and
  Lukin}}]{Duan2003}
\bibinfo{author}{\bibfnamefont{L.-M.} \bibnamefont{Duan}},
  \bibinfo{author}{\bibfnamefont{E.}~\bibnamefont{Demler}}, \bibnamefont{and}
  \bibinfo{author}{\bibfnamefont{M.~D.} \bibnamefont{Lukin}},
  \bibinfo{journal}{Phys.\ Rev.\ Lett.} \textbf{\bibinfo{volume}{91}},
  \bibinfo{pages}{090402} (\bibinfo{year}{2003}), ISSN \bibinfo{issn}{0031-9007
  (Print)}.

\bibitem[{\citenamefont{Bertoni et~al.}(2005)\citenamefont{Bertoni, Rontani, Goldoni, and Molinari}}]{Bertoni2005}
\bibinfo{author}{\bibfnamefont{A.}~\bibnamefont{Bertoni}},
  \bibinfo{author}{\bibfnamefont{M.}~\bibnamefont{Rontani}},
  \bibinfo{author}{\bibfnamefont{G.}~\bibnamefont{Goldoni}},
  \bibnamefont{and} \bibinfo{author}{\bibfnamefont{E.}~\bibnamefont{Molinari}},
  \bibinfo{journal}{Phys.\ Rev.\ Lett.} \textbf{\bibinfo{volume}{95}},
  \bibinfo{pages}{066806} (\bibinfo{year}{2005}).

\bibitem[{\citenamefont{Climente et~al.}(2006)\citenamefont{Climente, Bertoni, Rontani, Goldoni, and Molinari}}]{Climente2006}
\bibinfo{author}{\bibfnamefont{J.~I.}~\bibnamefont{Climente}},
  \bibinfo{author}{\bibfnamefont{A.}~\bibnamefont{Bertoni}},
  \bibinfo{author}{\bibfnamefont{M.}~\bibnamefont{Rontani}},
  \bibinfo{author}{\bibfnamefont{G.}~\bibnamefont{Goldoni}},
  \bibnamefont{and} \bibinfo{author}{\bibfnamefont{E.}~\bibnamefont{Molinari}},
  \bibinfo{journal}{Phys. Rev. B} \textbf{\bibinfo{volume}{74}},
  \bibinfo{pages}{125303} (\bibinfo{year}{2006}).

\bibitem[{\citenamefont{Climente et~al.}(2007)\citenamefont{Climente, Bertoni, Goldoni, Rontani, and Molinari}}]{Climente2007}
\bibinfo{author}{\bibfnamefont{J.~I.}~\bibnamefont{Climente}},
  \bibinfo{author}{\bibfnamefont{A.}~\bibnamefont{Bertoni}},
  \bibinfo{author}{\bibfnamefont{G.}~\bibnamefont{Goldoni}},
  \bibinfo{author}{\bibfnamefont{M.}~\bibnamefont{Rontani}},
  \bibnamefont{and} \bibinfo{author}{\bibfnamefont{E.}~\bibnamefont{Molinari}},
  \bibinfo{journal}{Phys. Rev. B} \textbf{\bibinfo{volume}{76}},
  \bibinfo{pages}{085305} (\bibinfo{year}{2007}).

\bibitem[{\citenamefont{Governale}(2002)}]{Governale2002}
\bibinfo{author}{\bibfnamefont{M.}~\bibnamefont{Governale}},
  \bibinfo{journal}{Phys. Rev. Lett.} \textbf{\bibinfo{volume}{89}},
  \bibinfo{pages}{206802} (\bibinfo{year}{2002}).

\bibitem[{\citenamefont{Kumada et~al.}(2006)\citenamefont{Kumada, Muraki, and
  Hirayama}}]{Kumada2006}
\bibinfo{author}{\bibfnamefont{N.}~\bibnamefont{Kumada}},
  \bibinfo{author}{\bibfnamefont{K.}~\bibnamefont{Muraki}}, \bibnamefont{and}
  \bibinfo{author}{\bibfnamefont{Y.}~\bibnamefont{Hirayama}},
  \bibinfo{journal}{Science} \textbf{\bibinfo{volume}{313}},
  \bibinfo{pages}{329} (\bibinfo{year}{2006}).

\bibitem[{\citenamefont{Nishi et~al.}(2006)\citenamefont{Nishi, Maksym, Austing, Hatano, Kouwenhoven, Aoki, Tarucha}}]{Nishi2006}
\bibinfo{author}{\bibfnamefont{Y.} \bibnamefont{Nishi}},
  \bibinfo{author}{\bibfnamefont{P.~A.}~\bibnamefont{Maksym}},
  \bibinfo{author}{\bibfnamefont{D.~G.}~\bibnamefont{Austing}},
  \bibinfo{author}{\bibfnamefont{T.}~\bibnamefont{Hatano}},
  \bibinfo{author}{\bibfnamefont{L.~P.}~\bibnamefont{Kouwenhoven}},
  \bibinfo{author}{\bibfnamefont{H.}~\bibnamefont{Aoki}},
  \bibnamefont{and} \bibinfo{author}{\bibfnamefont{S.}~\bibnamefont{Tarucha}},
  \bibinfo{journal}{Phys. Rev. B} \textbf{\bibinfo{volume}{74}},
  \bibinfo{pages}{033306} (\bibinfo{year}{2006}).

\bibitem[{\citenamefont{Ciorga et~al.}(2003)\citenamefont{Ciorga, Korkusinski, Pioro-Ladriere, Zawadzki, Hawrylak, Sachrajda}}]{Ciorga2003}
\bibinfo{author}{\bibfnamefont{M.} \bibnamefont{Ciorga}},
  \bibinfo{author}{\bibfnamefont{M.}~\bibnamefont{Korkusinski}},
  \bibinfo{author}{\bibfnamefont{M.}~\bibnamefont{Pioro-Ladriere}},
  \bibinfo{author}{\bibfnamefont{P.}~\bibnamefont{Zawadzki}},
  \bibinfo{author}{\bibfnamefont{P.}~\bibnamefont{Hawrylak}},
  \bibnamefont{and} \bibinfo{author}{\bibfnamefont{A.~S.}~\bibnamefont{Sachrajda}},
  \bibinfo{journal}{Phys. Stat. Sol. } \textbf{\bibinfo{volume}{238}},
  \bibinfo{pages}{325} (\bibinfo{year}{2003}).

\bibitem[{\citenamefont{Sachrajda et~al.}(2004)\citenamefont{Sachrajda,
  Korkusinski, Hawrylak, Ciorga, Pioro-Ladriere, and Zawadzki}}]{Sachrajda2004}
\bibinfo{author}{\bibfnamefont{A.~S.} \bibnamefont{Sachrajda}},
  \bibinfo{author}{\bibfnamefont{M.}~\bibnamefont{Korkusinski}},
  \bibinfo{author}{\bibfnamefont{P.}~\bibnamefont{Hawrylak}},
  \bibinfo{author}{\bibfnamefont{M.}~\bibnamefont{Ciorga}},
  \bibinfo{author}{\bibfnamefont{M.}~\bibnamefont{Pioro-Ladriere}},
  \bibnamefont{and} \bibinfo{author}{\bibfnamefont{P.}~\bibnamefont{Zawadzki}},
  \bibinfo{journal}{J.\ Magn.\ Magn.\ Mater.} \textbf{\bibinfo{volume}{272}},
  \bibinfo{pages}{E1273} (\bibinfo{year}{2004}).

\bibitem[{\citenamefont{Khandelwal et~al.}(2001)\citenamefont{Khandelwal, Dementyev, Kuzma, Barrett, Pfeiffer, and West}}]{Khandelwal2001}
\bibinfo{author}{\bibfnamefont{P.}~\bibnamefont{Khandelwal}},
  \bibinfo{author}{\bibfnamefont{A.~E.}~\bibnamefont{Dementyev}},
  \bibinfo{author}{\bibfnamefont{N.~N.}~\bibnamefont{Kuzma}},
  \bibinfo{author}{\bibfnamefont{S.~E.}~\bibnamefont{Barrett}},
  \bibinfo{author}{\bibfnamefont{L.~N.}~\bibnamefont{Pfeiffer}},
  \bibnamefont{and}
  \bibinfo{author}{\bibfnamefont{K.~W.}~\bibnamefont{West}},
  \bibinfo{journal}{Phys.\ Rev.\ Lett.} \textbf{\bibinfo{volume}{86}},
  \bibinfo{pages}{5353} (\bibinfo{year}{2001}).

\bibitem[{\citenamefont{Gervais et~al.}(2005)\citenamefont{Gervais, Stormer, Tsui, Kuhns, Moulton, Reyes, Pfeiffer, Baldwin, and West}}]{Gervais2005}
\bibinfo{author}{\bibfnamefont{G.}~\bibnamefont{Gervais}},
  \bibinfo{author}{\bibfnamefont{H.~L.}~\bibnamefont{Stormer}},
  \bibinfo{author}{\bibfnamefont{D.~C.}~\bibnamefont{Tsui}},
  \bibinfo{author}{\bibfnamefont{P.~L.}~\bibnamefont{Kuhns}},
  \bibinfo{author}{\bibfnamefont{W.~G.}~\bibnamefont{Moulton}},
  \bibinfo{author}{\bibfnamefont{A.~P.}~\bibnamefont{Reyes}},
  \bibinfo{author}{\bibfnamefont{L.~N.}~\bibnamefont{Pfeiffer}},
  \bibinfo{author}{\bibfnamefont{K.~W.}~\bibnamefont{Baldwin}},
  \bibnamefont{and}
  \bibinfo{author}{\bibfnamefont{K.~W.}~\bibnamefont{West}},
  \bibinfo{journal}{Phys.\ Rev.\ Lett.} \textbf{\bibinfo{volume}{94}},
  \bibinfo{pages}{196803} (\bibinfo{year}{2005}).

\bibitem[{\citenamefont{Hsieh et~al.}(2009)\citenamefont{Hsieh, Xia, Wray, Qian, Pal, Dil, Osterwalder, Meier, Bihlmayer, Kane, Hor, Cava, and Hasan}}]{Hsieh2009}
\bibinfo{author}{\bibfnamefont{D.}~\bibnamefont{Hsieh}},
  \bibinfo{author}{\bibfnamefont{Y.}~\bibnamefont{Xia}},
  \bibinfo{author}{\bibfnamefont{L.}~\bibnamefont{Wray}},
  \bibinfo{author}{\bibfnamefont{D.}~\bibnamefont{Qian}},
  \bibinfo{author}{\bibfnamefont{A.}~\bibnamefont{Pal}},
  \bibinfo{author}{\bibfnamefont{J.~H.}~\bibnamefont{Dil}},
  \bibinfo{author}{\bibfnamefont{J.}~\bibnamefont{Osterwalder}},
  \bibinfo{author}{\bibfnamefont{F.}~\bibnamefont{Meier}},
  \bibinfo{author}{\bibfnamefont{G.}~\bibnamefont{Bihlmayer}},
  \bibinfo{author}{\bibfnamefont{C.~L.}~\bibnamefont{Kane}},
  \bibinfo{author}{\bibfnamefont{Y.~S.}~\bibnamefont{Hor}},
  \bibinfo{author}{\bibfnamefont{R.~J.}~\bibnamefont{Cava}}, \bibnamefont{and}
  \bibinfo{author}{\bibfnamefont{M.~Z.}~\bibnamefont{Hasan}},
  \bibinfo{journal}{Science} \textbf{\bibinfo{volume}{323}},
  \bibinfo{pages}{919} (\bibinfo{year}{2009}).

\bibitem[{\citenamefont{M\"{u}hlbauer et~al.}(2009)\citenamefont{M\"{u}hlbauer, Binz, Jonietz, Pfleiderer, Rosch, Neubauer, Georgii, and B\"{o}ni}}]{Muehlbauer2009}
\bibinfo{author}{\bibfnamefont{S.}~\bibnamefont{M\"{u}hlbauer}},
  \bibinfo{author}{\bibfnamefont{B.}~\bibnamefont{Binz}},
  \bibinfo{author}{\bibfnamefont{F.}~\bibnamefont{Jonietz}},
  \bibinfo{author}{\bibfnamefont{C.}~\bibnamefont{Pfleiderer}},
  \bibinfo{author}{\bibfnamefont{A.}~\bibnamefont{Rosch}},
  \bibinfo{author}{\bibfnamefont{A.}~\bibnamefont{Neubauer}},
  \bibinfo{author}{\bibfnamefont{R.}~\bibnamefont{Georgii}}, \bibnamefont{and}
  \bibinfo{author}{\bibfnamefont{P.}~\bibnamefont{B\"{o}ni}},
  \bibinfo{journal}{Science} \textbf{\bibinfo{volume}{323}},
  \bibinfo{pages}{915} (\bibinfo{year}{2009}).

\bibitem[{\citenamefont{Saarikoski et~al.}(2004)\citenamefont{Saarikoski,
  Harju, Puska, and Nieminen}}]{Saarikoski2004}
\bibinfo{author}{\bibfnamefont{H.}~\bibnamefont{Saarikoski}},
  \bibinfo{author}{\bibfnamefont{A.}~\bibnamefont{Harju}},
  \bibinfo{author}{\bibfnamefont{M.~J.} \bibnamefont{Puska}}, \bibnamefont{and}
  \bibinfo{author}{\bibfnamefont{R.~M.} \bibnamefont{Nieminen}},
  \bibinfo{journal}{Phys.\ Rev.\ Lett.} \textbf{\bibinfo{volume}{93}},
  \bibinfo{pages}{116802} (\bibinfo{year}{2004}).

\bibitem[{\citenamefont{Tavernier, Anisimovas, and Peeters}(2004)}]{Tavernier2004}
  \bibinfo{author}{\bibfnamefont{M.~B.} \bibnamefont{Tavernier}},
  \bibinfo{author}{\bibfnamefont{E.}~\bibnamefont{Anisimovas}}, 
  \bibnamefont{and} 
  \bibinfo{author}{\bibfnamefont{F.~M.} \bibnamefont{Peeters}}
  \bibinfo{journal}{Phys.\ Rev.\ B} \textbf{\bibinfo{volume}{70}},
  \bibinfo{pages}{155321}  (\bibinfo{year}{2004}).

\bibitem[{\citenamefont{Saarikoski and Harju}(2005)}]{Saarikoski2005}
\bibinfo{author}{\bibfnamefont{H.}~\bibnamefont{Saarikoski}} \bibnamefont{and}
  \bibinfo{author}{\bibfnamefont{A.}~\bibnamefont{Harju}},
  \bibinfo{journal}{Phys.\ Rev.\ Lett.} \textbf{\bibinfo{volume}{94}},
  \bibinfo{pages}{246803} (\bibinfo{year}{2005}).

\bibitem[{\citenamefont{Yang et~al.}(2007)\citenamefont{Yang, Zhu, Dai, and
  Wang}}]{Yang2007}
\bibinfo{author}{\bibfnamefont{N.}~\bibnamefont{Yang}},
  \bibinfo{author}{\bibfnamefont{J.~L.} \bibnamefont{Zhu}},
  \bibinfo{author}{\bibfnamefont{Z.}~\bibnamefont{Dai}}, \bibnamefont{and}
  \bibinfo{author}{\bibfnamefont{Y.}~\bibnamefont{Wang}}
  (\bibinfo{year}{2007}), \eprint{arXiv:cond-mat/0701766v1}.

\bibitem[{\citenamefont{Anisimovas, Tavernier, and Peeters}(2008)}]{Anisimovas2008}
  \bibinfo{author}{\bibfnamefont{E}~\bibnamefont{Anisimovas}}, 
  \bibinfo{author}{\bibfnamefont{M.~B.} \bibnamefont{Tavernier}},
  \bibnamefont{and} 
  \bibinfo{author}{\bibfnamefont{F.~M.} \bibnamefont{Peeters}}
  \bibinfo{journal}{Phys.\ Rev.\ B} \textbf{\bibinfo{volume}{77}},
  \bibinfo{pages}{045327}  (\bibinfo{year}{2008}).

\bibitem[{\citenamefont{Yang et~al.}(2005)\citenamefont{Yang, Hwang, and
  Park}}]{Yang2005}
\bibinfo{author}{\bibfnamefont{S.-R.~Eric}~\bibnamefont{Yang}},
  \bibinfo{author}{\bibfnamefont{N.~Y.} \bibnamefont{Hwang}}, \bibnamefont{and}
  \bibinfo{author}{\bibfnamefont{S.}~\bibnamefont{Park}},
  \bibinfo{journal}{Phys.\ Rev.\ B} \textbf{\bibinfo{volume}{72}},
  \bibinfo{pages}{165337}  (\bibinfo{year}{2005}).

\bibitem[{\citenamefont{Petkovic and Milovanovic}(2007)}]{Petkovic2007}
  \bibinfo{author}{\bibfnamefont{A.}~\bibnamefont{Petkovic}} \bibnamefont{and}
  \bibinfo{author}{\bibfnamefont{M.~V.} \bibnamefont{Milovanovic}},
  \bibinfo{journal}{Phys.\ Rev.\ Lett.} \textbf{\bibinfo{volume}{98}},
  \bibinfo{pages}{066808} (\bibinfo{year}{2007}).

\bibitem[{\citenamefont{Milovanovic, Dobardzic, and Radovic}(2009)}]{Milovanovic2009}
  \bibinfo{author}{\bibfnamefont{M.~V.} \bibnamefont{Milovanovic}},
  \bibinfo{author}{\bibfnamefont{E.}~\bibnamefont{Dobardzic}}, 
  \bibnamefont{and} 
  \bibinfo{author}{\bibfnamefont{Z.}~\bibnamefont{Radovic}},
  \bibinfo{journal}{Phys. Rev. B} \textbf{\bibinfo{volume}{80}},
  \bibinfo{pages}{125305} (\bibinfo{year}{2009}).

\bibitem[{\citenamefont{Moon et~al.}(1995)\citenamefont{Moon, Mori, Yang,
  Girvin, MacDonald, Zheng, Yoshioka, and Zhang}}]{Moon1995}
\bibinfo{author}{\bibfnamefont{K.}~\bibnamefont{Moon}},
  \bibinfo{author}{\bibfnamefont{H.}~\bibnamefont{Mori}},
  \bibinfo{author}{\bibfnamefont{K.}~\bibnamefont{Yang}},
  \bibinfo{author}{\bibfnamefont{S.~M.}~\bibnamefont{Girvin}},
  \bibinfo{author}{\bibfnamefont{A.~H.}~\bibnamefont{MacDonald}},
  \bibinfo{author}{\bibfnamefont{L.}~\bibnamefont{Zheng}},
  \bibinfo{author}{\bibfnamefont{D.}~\bibnamefont{Yoshioka}}, \bibnamefont{and}
  \bibinfo{author}{\bibfnamefont{S.~C.}~\bibnamefont{Zhang}},
  \bibinfo{journal}{Phys.\ Rev.\ B} \textbf{\bibinfo{volume}{51}},
  \bibinfo{pages}{5138} (\bibinfo{year}{1995}).

\bibitem[{\citenamefont{Jacak et~al.}(1997)\citenamefont{Jacak, Hawrylak, and
  W{\'{o}}js}}]{jacak97:quant.dots}
\bibinfo{author}{\bibfnamefont{L.}~\bibnamefont{Jacak}},
  \bibinfo{author}{\bibfnamefont{P.}~\bibnamefont{Hawrylak}}, \bibnamefont{and}
  \bibinfo{author}{\bibfnamefont{A.}~\bibnamefont{W{\'{o}}js}},
  \emph{\bibinfo{title}{Quantum Dots}} (\bibinfo{publisher}{Springer},
  \bibinfo{address}{Berlin}, \bibinfo{year}{1997}).

\bibitem[{orb()}]{orbitals}
\bibinfo{note}{These orbital wave functions are well known. However, various
  errors in sign and factors exist in several published versions. The
  expression shown in Eq.~(\ref{eq:phiNM}) is correct both in terms of the
  energy eigenvalue, and, importantly, in terms of relations such as $a^\dagger
  |n,m\rangle = \sqrt{n+1} |n+1,m\rangle$.}

\bibitem[{\citenamefont{Abramowitz and Stegun}(1965)}]{Abramowitz1965}
\bibinfo{author}{\bibfnamefont{M.}~\bibnamefont{Abramowitz}} \bibnamefont{and}
  \bibinfo{author}{\bibfnamefont{I.~A.} \bibnamefont{Stegun}},
  \emph{\bibinfo{title}{Handbook of mathematical functions: with formulas,
  graphs, and mathematical tables}} (\bibinfo{publisher}{Dover Publications},
  \bibinfo{address}{New York}, \bibinfo{year}{1965}).

\bibitem[{\citenamefont{Kyriakidis et~al.}(2002)\citenamefont{Kyriakidis,
  Pioro-Ladriere, Ciorga, Sachrajda, and
  Hawrylak}}]{Kyriakidis02:Voltage-tunable-singlet-triplet}
\bibinfo{author}{\bibfnamefont{J.}~\bibnamefont{Kyriakidis}},
  \bibinfo{author}{\bibfnamefont{M.}~\bibnamefont{Pioro-Ladriere}},
  \bibinfo{author}{\bibfnamefont{M.}~\bibnamefont{Ciorga}},
  \bibinfo{author}{\bibfnamefont{A.~S.} \bibnamefont{Sachrajda}},
  \bibnamefont{and} \bibinfo{author}{\bibfnamefont{P.}~\bibnamefont{Hawrylak}},
  \bibinfo{journal}{Phys. Rev. B} \textbf{\bibinfo{volume}{66}},
  \bibinfo{pages}{035320} (\bibinfo{year}{2002}).

\bibitem[{\citenamefont{Stevenson and Kyriakidis}(2011)}]
  {Stevenson11:Fractional-orbital}
\bibinfo{author}{\bibfnamefont{C.~J.} \bibnamefont{Stevenson}} \bibnamefont{and}
  \bibinfo{author}{\bibfnamefont{J.}~\bibnamefont{Kyriakidis}},
  \bibinfo{journal}{Can. J. Phys.}  \textbf{\bibinfo{volume}{89}},
  \bibinfo{pages}{213} (\bibinfo{year}{2011}).

\bibitem[{\citenamefont{Saarikoski et~al.}(2010)\citenamefont{Saarikoski,
      Reimann, Harju, and Manninen}}]{Saarikoski2010}
\bibinfo{author}{\bibfnamefont{H.}~\bibnamefont{Saarikoski}},
  \bibinfo{author}{\bibfnamefont{S.~M.} \bibnamefont{Reimann}},
  \bibinfo{author}{\bibfnamefont{A.}~\bibnamefont{Harju}},
  \bibnamefont{and}
  \bibinfo{author}{\bibfnamefont{M.}~\bibnamefont{Manninen}},
  \bibinfo{journal}{Rev. Mod. Phys.} \textbf{\bibinfo{volume}{82}},
  \bibinfo{pages}{2785} (\bibinfo{year}{2010}).

\bibitem[{\citenamefont{Loss and Braun}(1996)}]{loss:1996:berry_phase}
\bibinfo{author}{\bibfnamefont{H.-B.}~\bibnamefont{Braun}} \bibnamefont{and}
  \bibinfo{author}{\bibfnamefont{D.} \bibnamefont{Loss}},
  \bibinfo{journal}{Phys. Rev. B} \textbf{\bibinfo{volume}{53}},
  \bibinfo{pages}{3237} (\bibinfo{year}{1996}).

\bibitem[{\citenamefont{Stevenson and Kyriakidis}(2010)}]{Stevenson:Unpublished}
\bibinfo{author}{\bibfnamefont{C.~J.}~\bibnamefont{Stevenson}}
  \bibnamefont{and}
  \bibinfo{author}{\bibfnamefont{J.}~\bibnamefont{Kyriakidis}} 
  \bibinfo{note}{(unpublished)}.

\bibitem[{\citenamefont{Hanson et~al.}(2005)\citenamefont{Hanson, van Beveren, Vink, Elzerman, Naber, Koppens, Kouwenhoven, and Vandersypen}}]{Hanson2005}
\bibinfo{author}{\bibfnamefont{R.}~\bibnamefont{Hanson}},
  \bibinfo{author}{\bibfnamefont{L.~H.~Willems} \bibnamefont{van~Beveren}},
  \bibinfo{author}{\bibfnamefont{I.~T.}~\bibnamefont{Vink}},
  \bibinfo{author}{\bibfnamefont{J.~M.} \bibnamefont{Elzerman}},
  \bibinfo{author}{\bibfnamefont{W.~J.~M.} \bibnamefont{Naber}},
  \bibinfo{author}{\bibfnamefont{F.~H.~L.} \bibnamefont{Koppens}},
  \bibinfo{author}{\bibfnamefont{L.~P.} \bibnamefont{Kouwenhoven}},
  \bibnamefont{and}
  \bibinfo{author}{\bibfnamefont{L.~M.~K.}~\bibnamefont{Vandersypen}},
  \bibinfo{journal}{Phys. Rev. Lett.} \textbf{\bibinfo{volume}{94}},
  \bibinfo{pages}{196802} (\bibinfo{year}{2005}).

\bibitem[{\citenamefont{Barthel et~al.}(2009)\citenamefont{Barthel, Reilly, Marcus, Hanson, and Gossard}}]{Barthel2009}
\bibinfo{author}{\bibfnamefont{C.}~\bibnamefont{Barthel}},
  \bibinfo{author}{\bibfnamefont{D.~J.} \bibnamefont{Reilly}},
  \bibinfo{author}{\bibfnamefont{C.~M.}~\bibnamefont{Marcus}},
  \bibinfo{author}{\bibfnamefont{M.~P.}~\bibnamefont{Hanson}},
  \bibnamefont{and}
  \bibinfo{author}{\bibfnamefont{A.~C.}~\bibnamefont{Gossard}},
  \bibinfo{journal}{Phys. Rev. Lett.} \textbf{\bibinfo{volume}{103}},
  \bibinfo{pages}{160503} (\bibinfo{year}{2009}).

\bibitem[{\citenamefont{Negele and Orland}(1988)}]{Negele1988}
\bibinfo{author}{\bibfnamefont{J.~W.} \bibnamefont{Negele}} \bibnamefont{and}
  \bibinfo{author}{\bibfnamefont{H.}~\bibnamefont{Orland}},
  \emph{\bibinfo{title}{Quantum many-particle systems}}
  (\bibinfo{publisher}{Addison-Wesley}, \bibinfo{year}{1988}).

\end{thebibliography}

\end{document}